\documentclass[iop, apj, tighten, numberedappendix]{emulateapj}
\slugcomment{{To appear in The Astrophysical Journal.}}

\usepackage{amsmath,amsthm}
\usepackage{amssymb}

\usepackage{mathtools}
\usepackage{empheq}
\usepackage{bbm}
\usepackage{bm}

\usepackage{hyperref}

\usepackage{epstopdf}

\usepackage{color} 

\bibpunct{(}{)}{;}{a}{}{,}

\newcommand{\myemail}{vboening@leibniz-kis.de}

\newcommand\id{{\mathrm d}}

\newcommand\Cref{C_{\text{ref}}}
\newcommand\cref{\Cref}

\newcommand\MF{\mathrm{MF}}
\newcommand\XT{\mathrm{XT}}
\newcommand\err{\mathrm{error}}
\newcommand\FWHM{\mathrm{FWHM}}

\newcommand\cK{\mathcal{K}}

\DeclareMathOperator{\Cov}{Cov}

\newcommand\mpersec{\rm{m}\,\rm{s}^{-1}}

\renewcommand\apjl{ApJL}
\renewcommand\solphys{SoPh}

\renewcommand\cite\citet

\shorttitle{Inversions for Meridional Flow using Spherical Born Kernels}

\shortauthors{B\"oning et al.}

\begin{document}


\title{Inversions for Deep Solar Meridional Flow Using Spherical Born Kernels}

\author{Vincent~G.~A.~B\"oning}
\affil{Kiepenheuer-Institut f\"ur Sonnenphysik, 79104 Freiburg, Germany}
\email{\myemail}
\author{Markus~Roth}
\affil{Kiepenheuer-Institut f\"ur Sonnenphysik, 79104 Freiburg, Germany}
\author{Jason Jackiewicz}
\affil{New Mexico State University, Las Cruces, NM 88001, USA}
\author{Shukur Kholikov}
\affil{National Solar Observatory, Tucson, AZ 85719, USA}


\begin{abstract}

The solar meridional flow is a crucial ingredient in modern dynamo theory. Seismic estimates of this flow have, however, been contradictory in deeper layers below about $0.9\,R_\sun$. 
Results from time-distance helioseismology have so far been obtained using the ray approximation. Here, we perform inversions using the Born approximation.
The initial result is similar to the result previously obtained by \cite{Jackiewicz2015} using ray kernels while using the same set of GONG data and the SOLA inversion technique. However, we show that the assumption of uncorrelated measurements used in earlier studies may lead to inversion errors being underestimated by a factor of about two to four. 
In a second step, refined inversions are performed using the full covariance matrix and a regularization for cross-talk. 
As the results are found to depend on the threshold used in the singular value decomposition, they were obtained for a medium threshold ($10^{-7}-10^{-5}$, about 50\% of the values used) and a threshold lower by a factor of 10 (about 70\% of the values used).
The result obtained with the medium threshold is again similar to the original, with less latitudinal variation. However, using the lower threshold, the inverted flow in the southern hemisphere shows two or three cells stacked radially depending on the associated radial flows. 
Both the single-cell and the multi-cell profiles are consistent with the measured travel times. 
All our results confirm a shallow return flow at about $0.9\,R_\sun$.

\end{abstract}




\section{Introduction}                                
\label{secintro}

Inferring the structure of the meridional flow in the deep solar interior has attracted considerable attention in recent years. While a conclusion on the flows in near-surface regions seems to have been reached \citep[e.g.,][]{GB2005,Miesch2005LRSP}, the most recent measurements of the deep meridional flow \citep[e.g.,][]{Hathaway2012,Zhao2013,Schad2013,Jackiewicz2015,Rajaguru2015} give seemingly contradictory results in deeper layers below about $0.9\,R_\sun$, favoring a single-cell or multi-cell picture of the flow as summarized in the introduction of \cite{Boening2017}.

Possible reasons for this discrepancy may include systematic effects like a center-to-limb effect in time-distance helioseismology \citep[e.g.,][]{Zhao2012CtoL,Zhao2016,Liang2017}, perturbation of solar mode eigenfunctions by convection \citep[e.g.,][]{Baldner2012}, systematic effects introduced by magnetic fields \citep[e.g.,][]{Liang2015SurfaceMagnOnMeridFlow,Liang2015ProbingMagneticFields}, B- or P-angle variations \citep[e.g.,][]{Kholikov2014,Liang2017}, as well as differences in the instruments used and the time period considered.

In addition, there are several efforts underway to develop or validate new methods for inferring the meridional flow, in particular using local helioseismic techniques \citep[e.g.,][]{Boening2016,Roth2016,Gizon2017}.

As inversions for the meridional flow with time-distance helioseismology \citep{Zhao2013,Jackiewicz2015,Rajaguru2015} have so far been modeled using the rather classical ray approximation \citep{KosovichevDuvall1997}, the Born approximation has been brought forward very recently as an alternative \citep{Boening2016,Gizon2017}. Instead of assuming that travel times of acoustic waves can be perturbed by a flow along a ray path as in the ray approximation, the Born approximation \citep[e.g.][]{GB2002} assumes that the whole wave field is scattered by the flow. Therefore, the travel time of a wave packet can be perturbed by a flow distant from the ray path. The Born approximation has been well tested and validated in Cartesian geometry for inferring small-scale flows \citep[e.g.,][]{Svanda2011,Svanda2013c,Svanda2013a,DeGrave2014QS,DeGrave2014Sunspot,DeGrave2015,Fournier2016}. As it is generally thought of as a more accurate model of the physics in the solar interior \citep[e.g.,][]{Birch2001,Birch2004,Couvidat2006,BG2007}, it is a method worth exploring for inferring large-scale flows such as the meridional flow.

Born approximation sensitivity functions (kernels) have very recently been validated in time-distance helioseismology of the meridional flow \citep{Boening2017}. In this study, we will perform inversions for the deep meridional flow with these spherical Born kernels. As was shown by \cite{Boening2016}, phase-speed filtered measurements seem particularly useful for this endeavor. We will therefore use the phase-speed filtered travel-time measurements obtained by \cite{Kholikov2014} in this work. These measurements have been inverted for the meridional flow by \cite{Jackiewicz2015} using the SOLA method \citep{Pijpers1994,Jackiewicz2012}.

The first objective of this work is to answer the question whether the inversion results inferred by \cite{Jackiewicz2015} using ray kernels can be confirmed using Born kernels or whether a different conclusion may be reached. Our second objective is to study in detail different sources of systematic errors in the inversion process, such as the error propagation and the cross-talk of the radial flow into the inversion for the horizontal flow component. For this work, we will employ an analytic formula for the covariance of travel-time measurements (\citealp{GB2004}, \citealp{Fournier2014}). We will also use different strategies for analyzing the cross-talk and other systematics in SOLA inversions using Born kernels that were also employed by various authors in Cartesian inversions for small-scale flows \citep[e.g.,][]{Svanda2011,DeGrave2014QS,Fournier2016}.

The paper is organized as follows. Section~\ref{secdatainv} provides a short introduction to the data set and the SOLA inversion technique used. Section~\ref{secborn} gives a summary of the computation of spherical Born kernels, which are used to model the travel-time measurements in this work. An analytic formula from the literature is used in Section~\ref{seccov} to obtain a model for the covariance matrix of the measurements. A comparison of an initial inversion for the meridional flow using Born kernels to the result obtained by \cite{Jackiewicz2015} is presented in Section~\ref{secfirstcomp}. This inversion is performed under the assumption of uncorrelated measurements (i.e., a diagonal covariance matrix) and without including a regularization for cross-talk as it was done in previous inversions for the meridional flow using time-distance helioseismology \citep{Zhao2013,Jackiewicz2015,Rajaguru2015}. In this section, we also provide a detailed study on the error propagation as well as the cross-talk between flow components. This information is used in Section~\ref{secfullinv} to perform refined inversions for the meridional flow that include the full covariance matrix and a regularization for cross-talk. The results of this study are discussed in Section~\ref{secdiscussion} and conclusions are presented in Section~\ref{secconclusions}.

\section{Data and Inversion Technique}
\label{secdatainv}

In this paper, we use the travel-time measurements obtained and described by \cite{Kholikov2014}. The travel times were obtained using the GONG instrument and the data of 652 days between 2004 and 2012, where the duty cycle was high and the B-angle not too large. This period includes the declining phase of cycle 23 and the rising phase of cycle 24. An average sunspot number of 36 for the days used \citep{SSN_SILSO_2004_2012} indicates that the data are taken from times of moderate to low activity. North minus south (N -- S) point-to-arc travel times were obtained for a total of 72 travel distances ($\Delta$) and 384 latitude bins ($\lambda$). In addition, east minus west (E -- W) travel times were measured in order to correct for a systematic center-to-limb effect detected by earlier studies (e.g., \citealp{Zhao2012}, \citealp{Zhao2013}). The N -- S travel times are corrected for this systematic effect by subtracting the E -- W measurements.

The travel times were obtained using a Gabor fit \citep{KosovichevDuvall1997}. \cite{Jackiewicz2015} inverted this set of travel times for the horizontal component of the meridional flow using a standard SOLA technique \citep{Pijpers1994,Jackiewicz2012}. This inversion technique will be the starting point of the work performed in this study. Our goal is to find the meridional flow with horizontal component $v_\theta(r,\theta)$ and radial component $v_r(r,\theta)$ that satisfies for all travel-time measurements $\delta\tau_i$
\begin{align}
\delta \tau_i = & \int_0^{R_\sun}  \int_0^\pi  \, \sum_{k=r,\theta} K_k^i(r,\theta)  v_k(r,\theta) \; r \sin(\theta)\,\id\theta \,\id r \, \nonumber \\
& + \, \epsilon_i,
\label{eqkernelgoal}
\end{align}
where $K_k^i$ is the travel-time sensitivity function of the measurement $i=(\Delta,\lambda)$ to a flow in the direction $k\in\{r,\theta\}$, and $\epsilon_i$ are measurement errors. Two-dimensional integrals are evaluated over the whole radial and latitudinal domains throughout this work. The meridional flow is sought at a series of target locations $(r_T,\theta_T)$ as a linear combination of the measured travel times, e.g., for the horizontal flow component,
\begin{equation}
v_\theta^{\rm{inv}}(r_T,\theta_T) = \sum_i w_i(r_T,\theta_T) \, \delta\tau_i.
\end{equation}
The aim is to construct averaging kernels $\cK_\theta$,
\begin{align}
\cK_\theta(r,\theta;r_T,\theta_T) = \sum_i K_\theta^i(r,\theta)w_i(r_T,\theta_T),
\end{align}
which are sufficiently near a Gaussian target function $T(r,\theta;r_T,\theta_T)$ with a certain radial and horizontal full width at half maximum ($\FWHM_r,\FWHM_\theta$) centered at the target location. This is achieved by minimizing the misfit given by
\begin{align}
\MF  = \iint  \left| \cK_\theta - T \right| ^2 \; r \sin(\theta)\,\id\theta \,\id r.
\end{align}
The minimization is performed with an additional regularization parameter $\mu$ that balances misfit and errors,
\begin{align}
\err^2 =\sum_{i,j} w_i(r_T,\theta_T) \Lambda_i^j w_j(r_T,\theta_T),
\end{align}
where the covariance matrix $\Lambda$ is given by
\begin{align}
\Lambda_i^j = \Cov \left[ \epsilon_i ,\, \epsilon_j\right].
\end{align}
The final cost function
\begin{align}
\chi(w_i&(r_T,\theta_T);\mu)  = \MF + \mu \,\err^2 
\end{align}
is then minimized at every target location subject to the constraint
\begin{align}
\iint \cK_\theta(r,\theta;r_T,\theta_T) \; r \sin(\theta)\,\id\theta \,\id r = 1.
\end{align}
The resulting weights can be obtained using a matrix inversion \citep[e.g.,][]{Svanda2011,Jackiewicz2012,Jackiewicz2015}.

\section{Forward-modeling using Born Kernels}
\label{secborn}

The forward-modeling of the travel-time measurements in \cite{Jackiewicz2015} was performed using ray kernels \citep[e.g.,][]{KosovichevDuvall1997}. Instead, we will use the Born approximation \citep[e.g.][]{GB2002} to model the effect of the flow on the travel times in this work.

While the ray approximation assumes that the acoustic waves in the solar interior are sensitive to a flow field only along a certain ray path, the Born approximation models the scattering of the full wave field due to advection in first order \citep{GB2002,BG2007}. The method used in this work for computing Born kernels was developed by \cite{Boening2016} and further refined by \cite{Boening2017}. 

As the computation of the sensitivity kernels depends on an accurate match between model and data power spectra, the following free parameters in the model were adjusted \citep[see][]{DeGrave2014QS,Boening2017}. 
For harmonic degree $l<100$, mode frequencies and damping rates from the GONG ftp site\footnote{\url{ftp://gong2.nso.edu/TSERIES/v1f/}} were used, and for $l\geq 100$, they were provided by Sylvain Korzennik \citep{Korzennik2013ApJ}. 
For small harmonic degrees, the fitted mode widths are much smaller than the frequency resolution of a day-long time series. Therefore, small mode widths were increased to a minimum value 40\% larger than the resolution of a day-long time series. This value was empirically found to produce a good match between data and kernel power spectra. As in \cite{Boening2017}, the model power spectrum was further corrected by an $l$-dependent factor, which may be seen as an optical transfer function correcting instrumental effects; see \cite{BirchKo2004} and \cite{BG2007}. In addition, the source correlation time, a free parameter that models the sources in the model, is adjusted to obtain a match between the mean frequency of the data and model power spectra. Finally, the phase-speed filters used in the data analysis procedure of \cite{Kholikov2014} were applied to the model power spectrum.

\begin{figure*}%
\begin{center}

\includegraphics[height=0.27\textwidth]{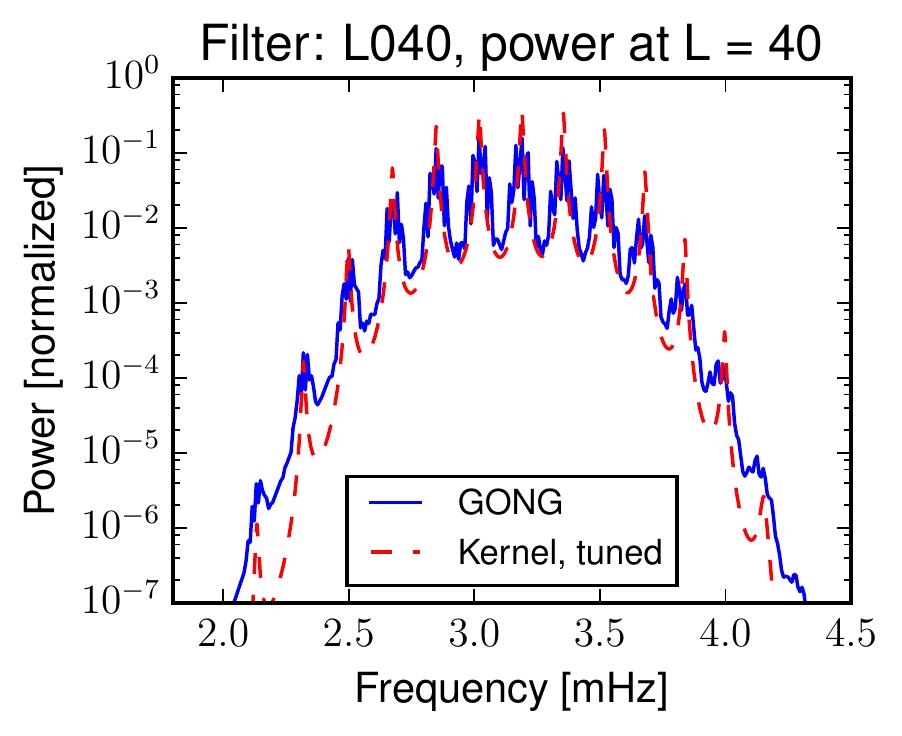}%
\includegraphics[height=0.27\textwidth]{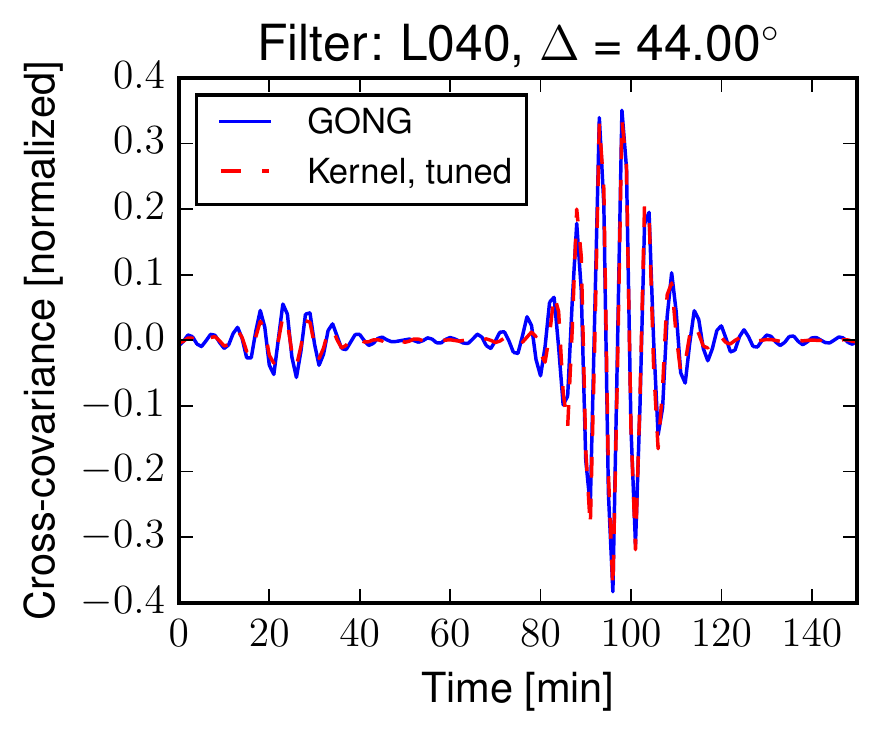}%
\includegraphics[height=0.27\textwidth]{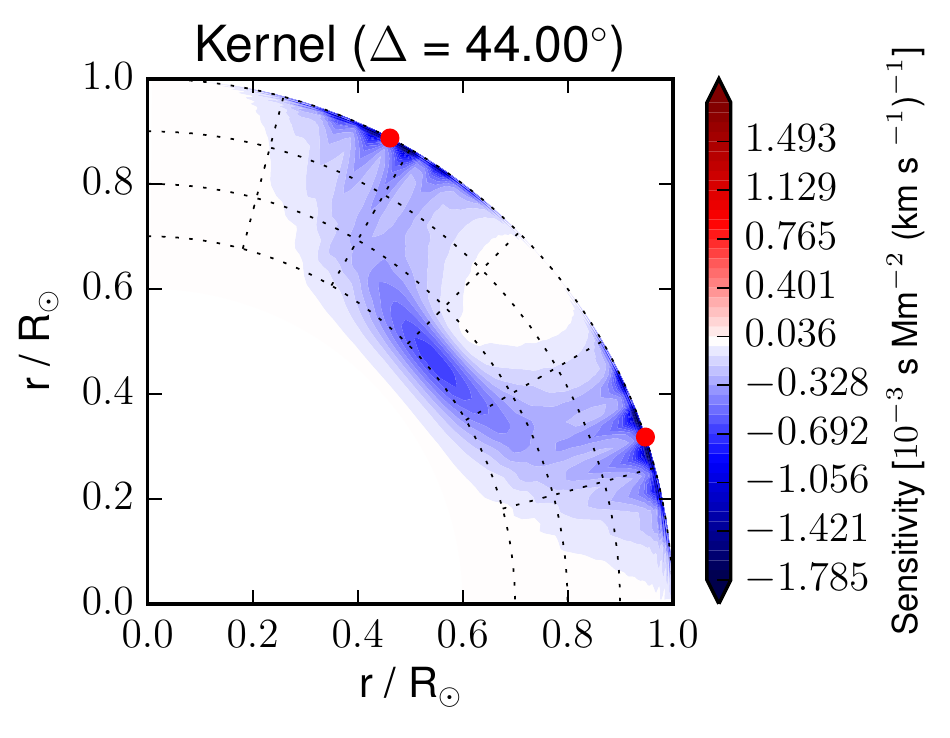}%

\includegraphics[height=0.27\textwidth]{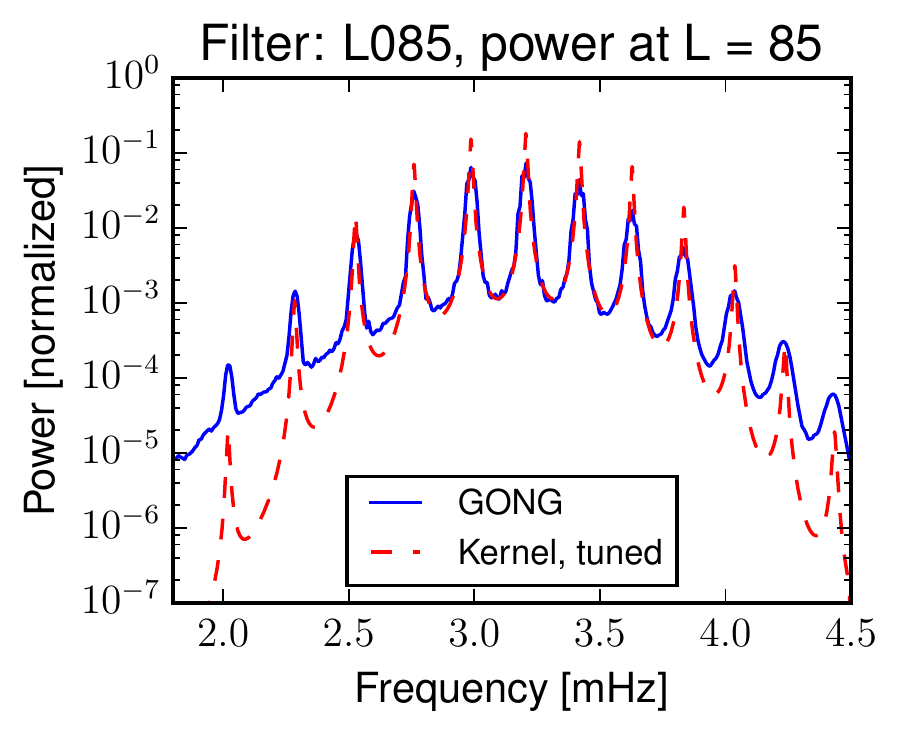}%
\includegraphics[height=0.27\textwidth]{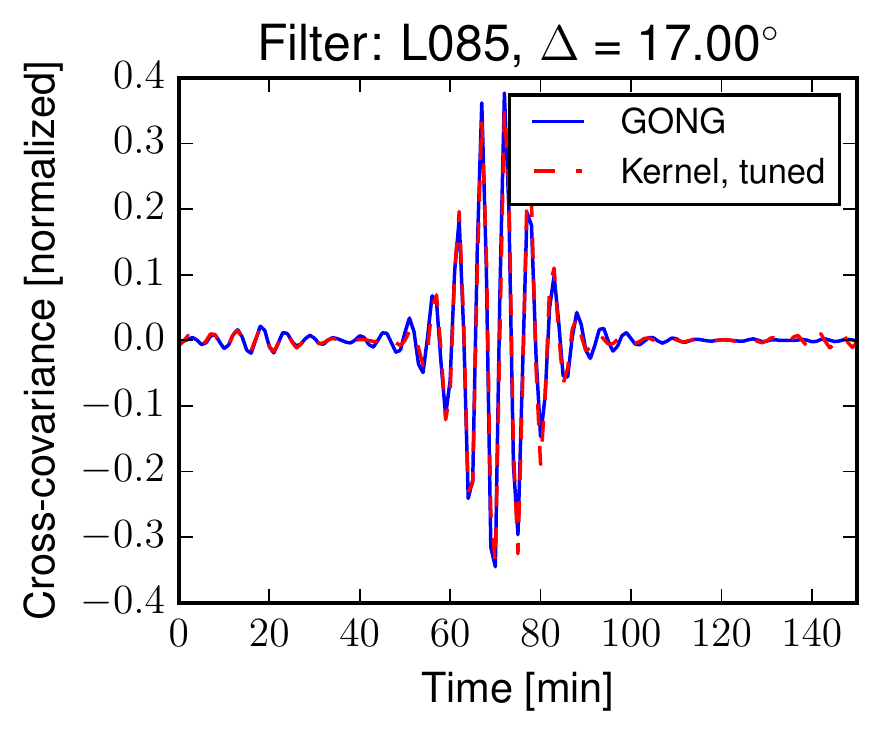}%
\includegraphics[height=0.27\textwidth]{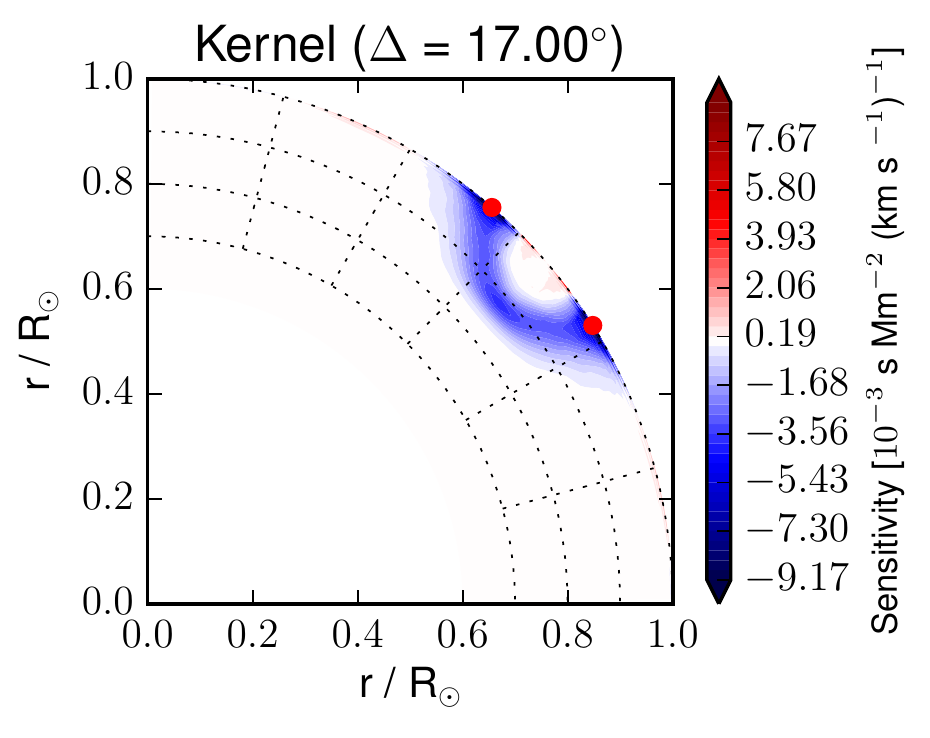}%

\end{center}

\caption{Key features of the model used to compute the Born kernels for this study. Shown are comparisons of model vs. GONG data for mean power spectra (left column) and cross-covariances (middle column), as well as an example kernel at a latitude of $41\degr$ (right column) for two example travel distances and corresponding filters (top and bottom rows). The kernels are saturated at 15 \% of their maximum value attained at the surface, and the locations of the observation points are marked with red dots.}%
\label{figborn}%
\end{figure*}

As a result, zero-order model power spectra and cross-covariances agree well with those obtained from GONG data; see Figure~\ref{figborn}. Only leakage effects visible in the GONG power spectrum at low harmonic degree (top-left panel) are not included in our model. However, no similar effect is visible in the cross-covariances (top middle panel), which are used to model the travel-time shift.

Inspection of kernel plots (right column in Figure~\ref{figborn}) also reveals similar results to \cite{Boening2017} with some changes introduced by differences in the filters and data power spectra.

For each of the 72 travel distances, one 3D kernel was computed. This kernel was then reprojected to the different latitudes and rotated to obtain the N -- S point-to-arc geometry with $30\degr$ wide arcs used in the data analysis procedure. After integrating the kernels along the azimuthal domain, they were smoothed and rebinned with respect to distance in a similar way to that in the data analysis with a final number of 45 distances. Finally, both the travel times and the kernels were further rebinned by a factor of two to obtain 192 latitude bins, which are used in the inversions performed in this work.

\section{Full Covariance Matrix}
\label{seccov}

In previous inversions for the deep meridional flow using time-distance helioseismology \citep{Zhao2013,Jackiewicz2015,Rajaguru2015}, the travel-time measurements were assumed to be uncorrelated, which is equivalent to a diagonal covariance matrix. In order to study the effect of this assumption, we will use both a diagonal and the full covariance matrix in this work.

Analytic formulas for computing the covariance of travel-time measurements have been proposed by \cite{GB2004} and \cite{Fournier2014}. They were obtained on the basis of an empirical model for the noise in the power spectrum, which assumes uncorrelated noise for different frequency bins, and were shown to be in good agreement with data and Monte Carlo simulations.

For the purpose of this work, we compute the covariance matrix using formula (B.6) or equivalently (13) from \citet[][see also Equation (28) in \citealp{GB2004}]{Fournier2014}. As suggested in \cite{Fournier2014}, for longer time series (one day in our case), only the first term in the formula has to be taken into account. As an ingredient, only the zero-order or mean cross-covariances for all combinations of measurements are needed. In this work, we use the mean power spectrum of the data to obtain mean cross-covariances for every possible distance needed in the formula. For computational reasons, we compute the covariance for point-to-point measurements only. 

The resulting diagonal entries of the covariance matrix should in principle be equal to the squared errors of the travel-time measurements used in this study. For two reasons this is not the case. Firstly, the analytic formula for the diagonal entries only depends on the travel distance and not on the latitude, in contrast to the errors of the measurements, which increase toward the limb. Second, the analytic values have a different scale compared to the squared errors as the measurements have been further averaged, e.g. over a period of 652 days. This is taken into account by renormalizing the result from the analytic formula using
\begin{align}
\tilde{\Lambda}_i^j =  \Lambda_i^j \frac{\sigma_i \, \sigma_j }{ \sqrt{\Lambda_i^i \Lambda_j^j} },
\end{align}
where $\sigma_i$ is the error of measurement $i$.

\begin{figure}
\includegraphics[width=\columnwidth]{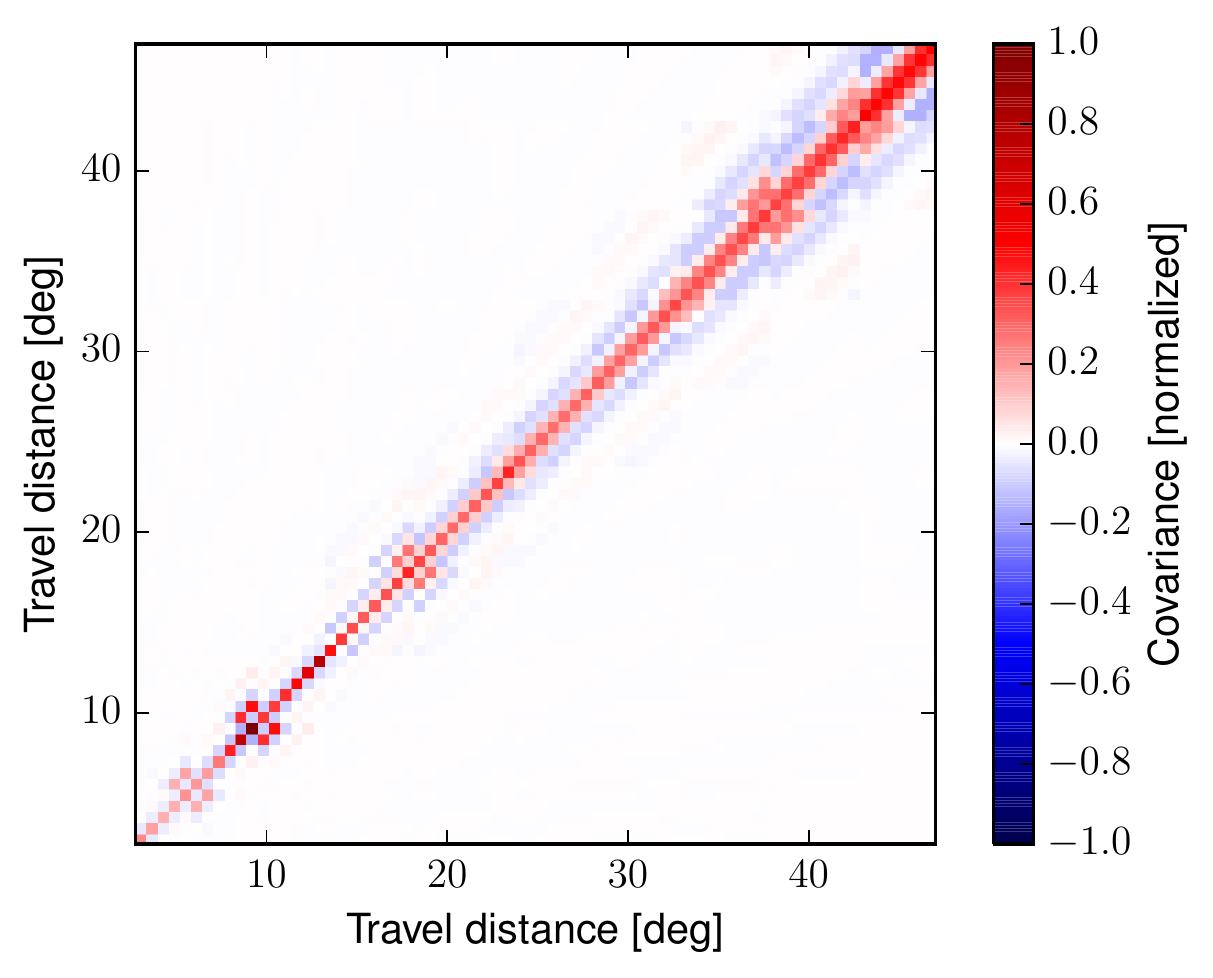}%
\caption{Cut through the covariance matrix $\Lambda_{\Delta,\lambda}^{\Delta',\lambda'}$ at $\lambda=\lambda'=0\degr$. Compare to \cite{Roth2007} for a similar matrix obtained from unfiltered MDI data. The minimal value in this plot is about $0.3$, similar to \cite{Roth2007}.}%
\label{figcov}
\end{figure}

A cut through the covariance matrix obtained is displayed in Figure~\ref{figcov}, which is in general agreement with the covariance matrix obtained by \cite{Roth2007} from unfiltered data.

\section{Horizontal SOLA Inversion: A First Comparison}
\label{secfirstcomp}

The aim of this section is to perform a first comparison of an inversion for the meridional flow using spherical Born kernels to an earlier inversion done with ray kernels. For this purpose, the inversion method is kept as close as possible to the one applied in \cite{Jackiewicz2015}. Specifically, we only invert for the horizontal component of the flow, we assume the covariance matrix to be diagonal, and we do not take the cross-talk from the radial flow component into the horizontal flow component into account when obtaining the inversion weights.

Our intention is to first answer the question as to whether and how the inversion result is affected by simply exchanging ray kernels with Born kernels. Second, we will compare the magnitude of the errors propagated with a diagonal and with the full covariance matrix, and we will analyze the magnitude of the cross-talk.

\subsection{Choice of Regularization Parameters}

The inversion procedure involves a number of free parameters, such as the full widths of the Gaussian target functions and the parameter $\mu$, which may vary with target location. We therefore first performed a number of test computations in order to get an overview over the parameter space involved. These tests showed that a scaling of the full widths of the target functions with the sound speed, as proposed in \cite{Pijpers1994}, is a reasonable choice. Near the surface ($r_T\geq0.95\,R_\sun$), this scaling would imply very small target widths. We therefore set minimum target widths of $\FWHM_{r,\rm{min}} = 0.03 R_\sun$ and $\FWHM_{\theta,\rm{min}} = 5 \degr$. This limitation is justified by the minimum scales of the kernels involved in this study as the minimum travel distance is around three degrees and the kernels extend a little farther than that. For the scaling of the full widths, we find values at the bottom of the convection zone at $r_T=0.7 R_\sun$ of $\FWHM_{r} = 0.09 R_\sun$ and $\FWHM_{\theta} = 20\degr$. At first sight, these values may seem very large, especially when compared with the averaging kernels obtained by \cite{Zhao2013} using the ray approximation, which are much more localized. However, taking a closer look at the spatial scales involved in the kernels shown in Figure~\ref{figborn}, which are much wider than ray kernels, the values obtained are plausible. Our test computations showed that, if the widths of the target functions are further decreased significantly, it is hard to obtain reasonable values for errors and misfit.

In addition, every inverse matrix in the inversion problem can be computed using different thresholds for the singular values (SVs). In practice, the inversion results do not show a large dependence on the choice of threshold in this inversion, where we use a diagonal covariance matrix.

In order to obtain an optimal choice for the remaining inversion parameter $\mu$, which controls the relation between errors and misfit in the inversion, we first do a number of inversions for a series of values for $\mu$ on a coarse grid of target locations. Finally, we choose an optimal value for $\mu$ for every target depth, using maximum threshold values for the misfit and the errors of the inverted flows. In practice, we find that it is often hard to obtain a reasonable inversion error without compromising the fit of the averaging kernel to the target kernel. We thus chose a maximum error of $1\,\mpersec$, similar to \cite{Jackiewicz2015}, and a maximum misfit of 0.2, when the misfit is normalized as 
\begin{align}
\MF_{\rm{norm}}  = \frac{\MF}{\iint  T ^2 \; r \sin(\theta)\,\id\theta \,\id r}.
\end{align}
We note here that it is possible to achieve much better values for the misfit, up to almost perfect agreement between the averaging and target kernels, but at an unacceptable expense in the errors. A maximum misfit of 0.2 is found to be about the largest possible value for obtaining an acceptable match between the averaging and target kernels. At the surface, this condition is relaxed by a factor of two, as we find that most of the misfit comes from locations just being a bit farther off, or a difficulty of the averaging kernel to achieve a Gaussian form. If no value for $\mu$ is found for the maximum error and misfit given, these conditions are relaxed step by step until possible inversion parameters are found. If several possible points are found, an optimal choice is made using the so-called L-curves \citep[e.g.,][]{Jackiewicz2015}. In the optimizing procedure, we consider a small number of target latitudes within $20\degr$ from the equator.

Using the coarse grid of target locations, optimal inversion parameters are chosen for every target depth of the spatial grid. For the final inversion, a finer target grid is defined, and for each location, the inversion parameters are interpolated between the values from the preparatory inversion.

\subsection{Inversion Results}

\begin{figure*}

\includegraphics[height=0.41\textwidth]{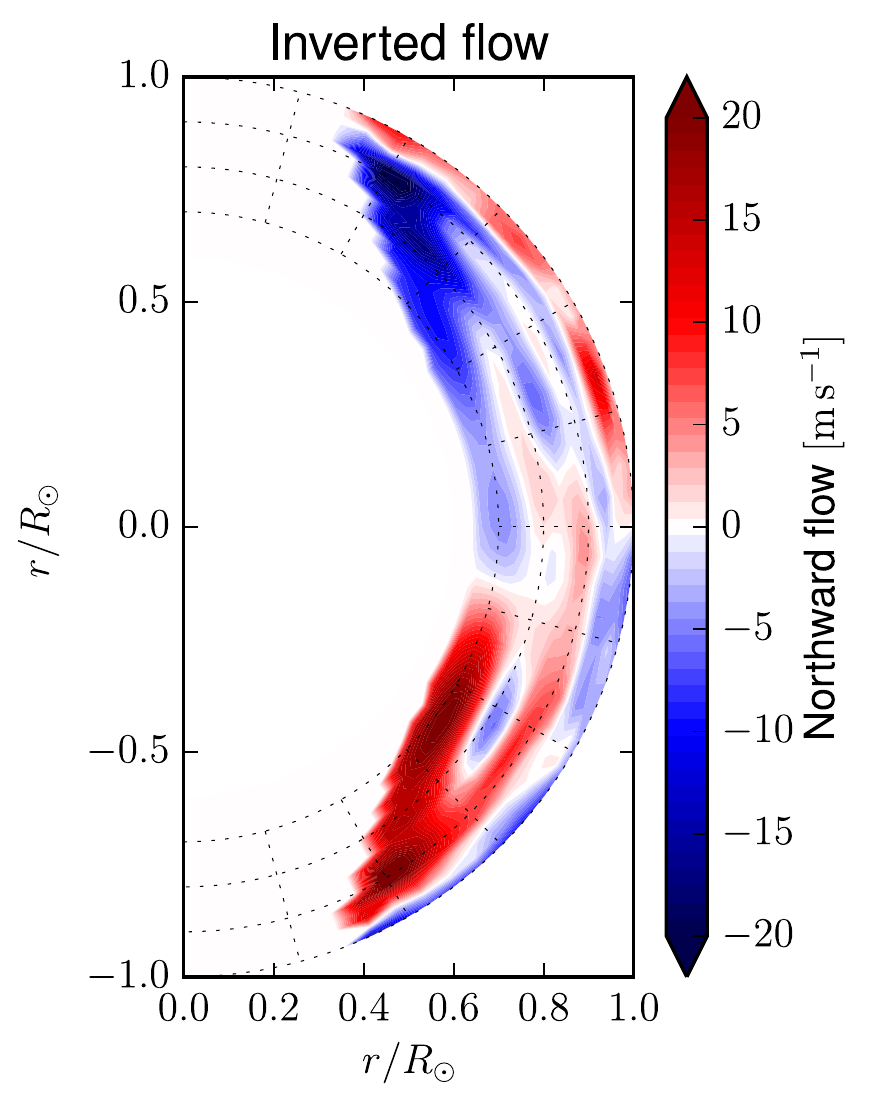}%
\includegraphics[height=0.41\textwidth]{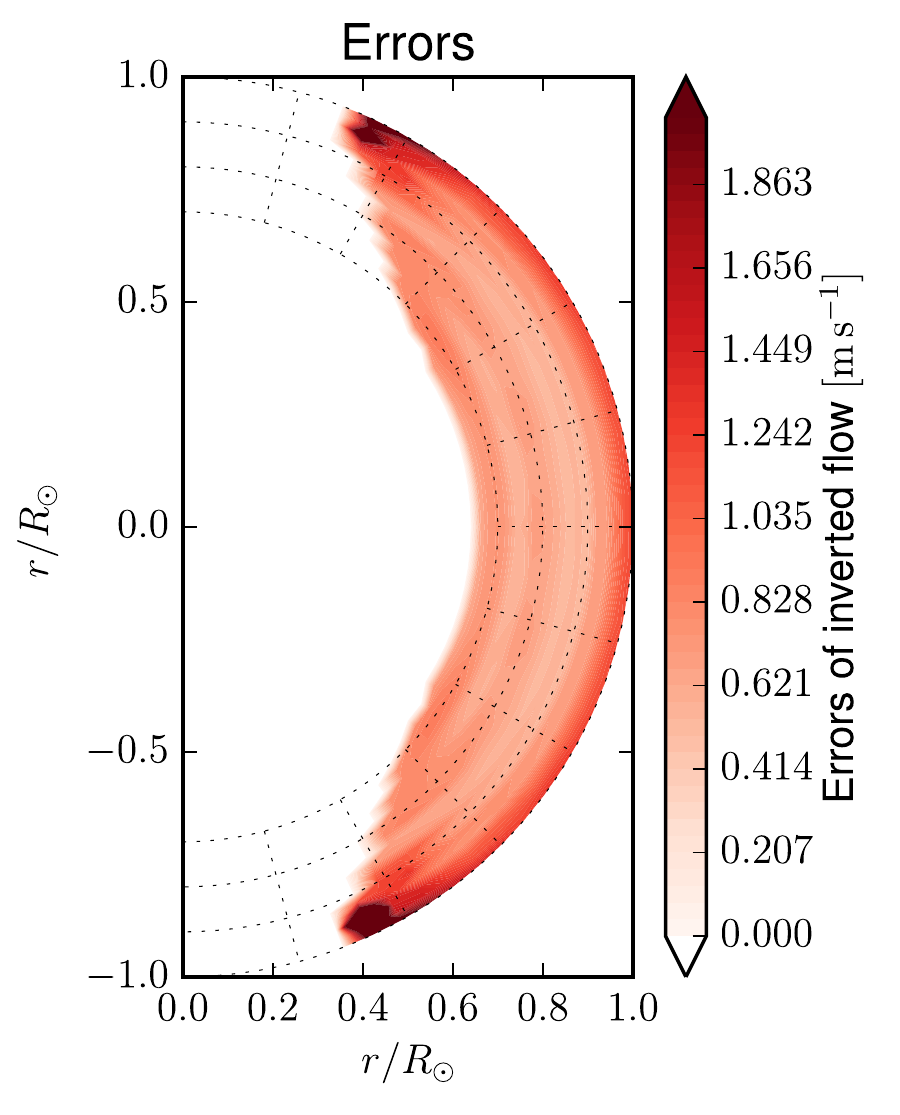}%
\includegraphics[height=0.41\textwidth]{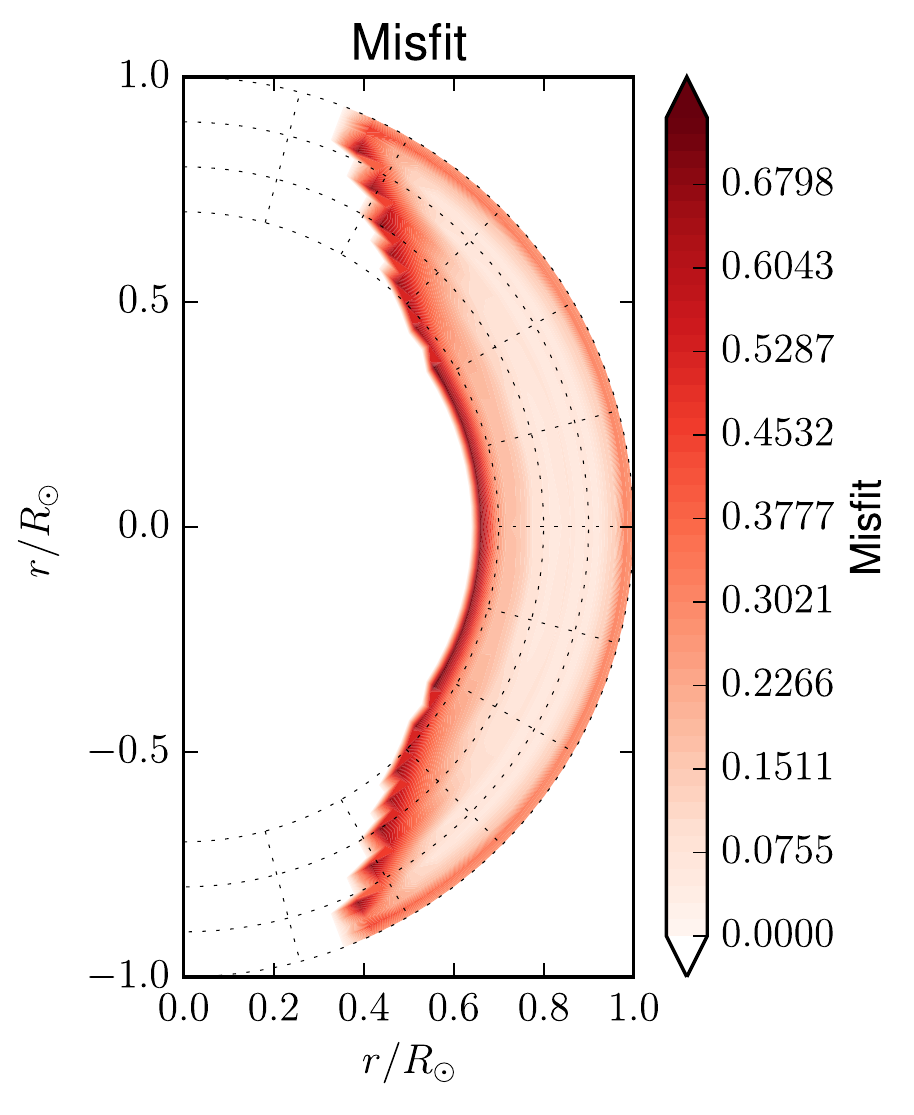}%

\caption{Initial result using GONG data and the SOLA inversion method from \cite{Jackiewicz2015}, assuming uncorrelated measurements. The left panel shows the inverted flow profile, the middle panel the corresponding errors (see also Figure~\ref{figerrscatter}), and the right panel the misfit. The error plot is saturated at the maximum values in the $\pm 40\degr$ latitude range.}
\label{figsolabornall}%
\end{figure*}

Inversion results are presented in Figure~\ref{figsolabornall}, where the inverted flow (left panel), the inversion errors (middle panel), and the misfit (right panel) are shown. We first note that the resulting meridional flow is similar to the inversion result obtained by \cite[][see Figure~4]{Jackiewicz2015} using ray kernels, at least qualitatively.

As this inversion result and the result obtained by \cite{Jackiewicz2015} are obtained using a diagonal covariance matrix, it is questionable whether the error estimate is correct. Using the inversion weights obtained in this inversion and the full covariance matrix computed as in Section~\ref{seccov}, it is possible to give more accurate estimates of the errors of the inverted flow and to estimate the impact of not taking the full covariance matrix into account.

\begin{figure}
\includegraphics[width=\columnwidth]{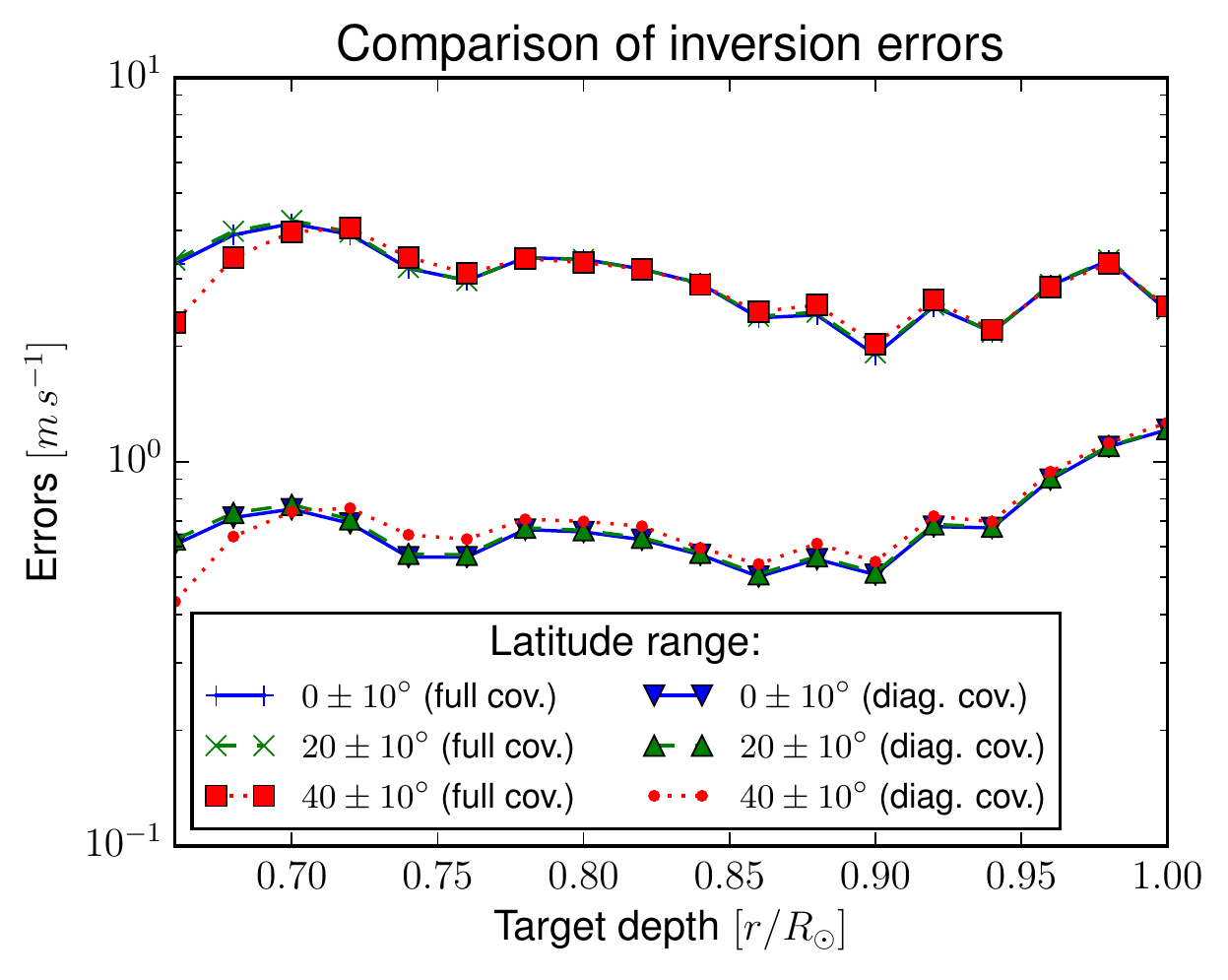}%
\caption{Errors for the inversion result presented in Figure~\ref{figsolabornall}. Errors computed using the full covariance matrix (lines near the top) can be seen to be about two to four times underestimated when the measurements are assumed to be uncorrelated (diagonal covariance matrix, lines near the bottom).}%
\label{figerrscatter}%
\end{figure}

In Figure~\ref{figerrscatter}, inversion errors using the diagonal covariance matrix (lines near the bottom) can be compared to errors obtained using the full covariance matrix (lines near the top) as a function of target depth for a series of target latitudes. Both errors are obtained using the same inversion weights. It can be seen that the errors from the diagonal covariance underestimate the errors from the full covariance by a factor of about two to four. For this comparison, the values on the diagonal of the two covariance matrices are identical prior to the final rebinning of the measurements by a factor of two in latitude; see Section~\ref{secdatainv}. If they were to be set equal after this final rebinning, the errors from the diagonal covariance would increase on average by about 35 \% and they would still underestimate the errors from the full covariance by a factor of 1.5 - 3.

\begin{figure*}%

\includegraphics[width=0.5\textwidth]{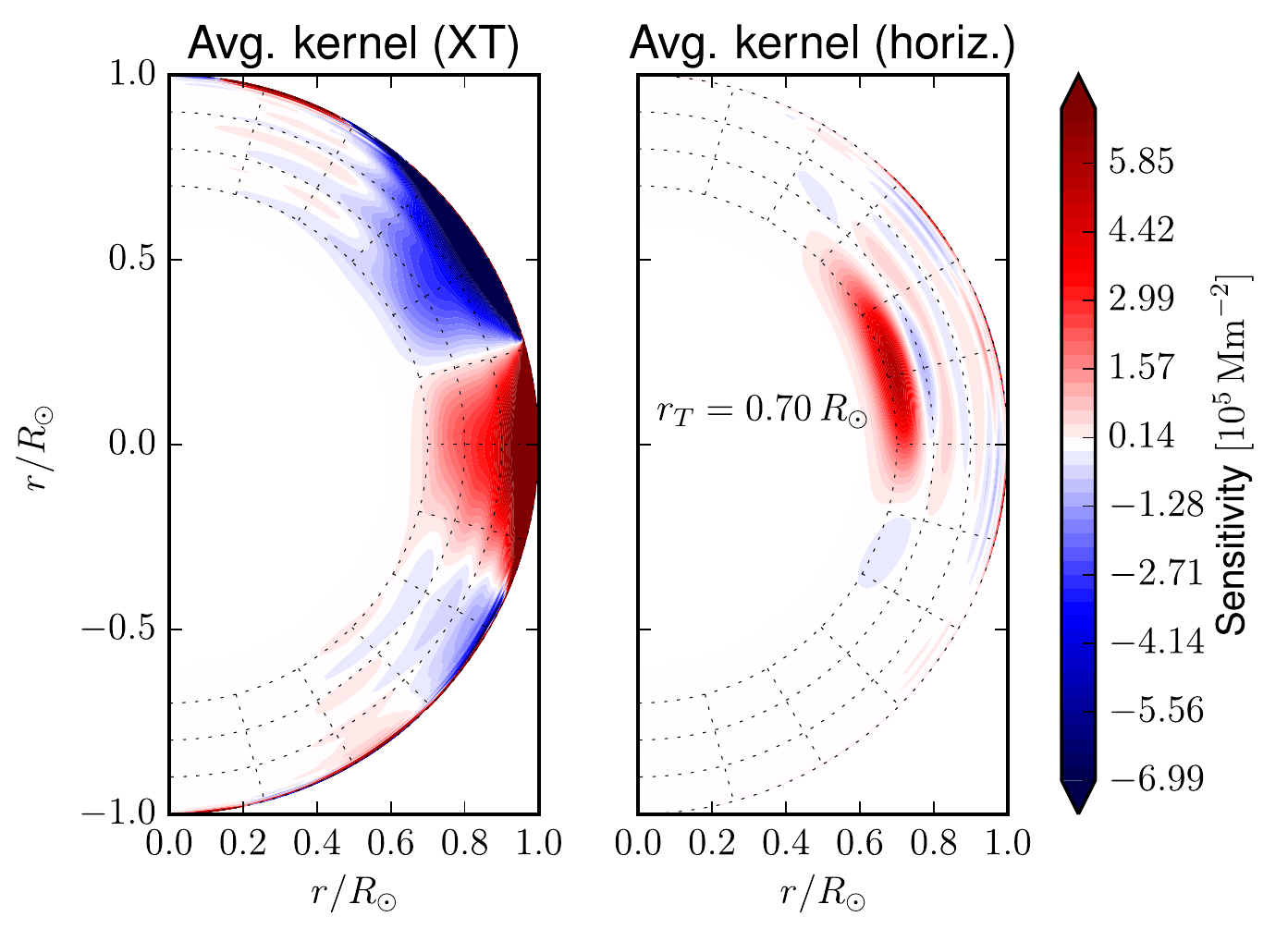}%
\includegraphics[width=0.5\textwidth]{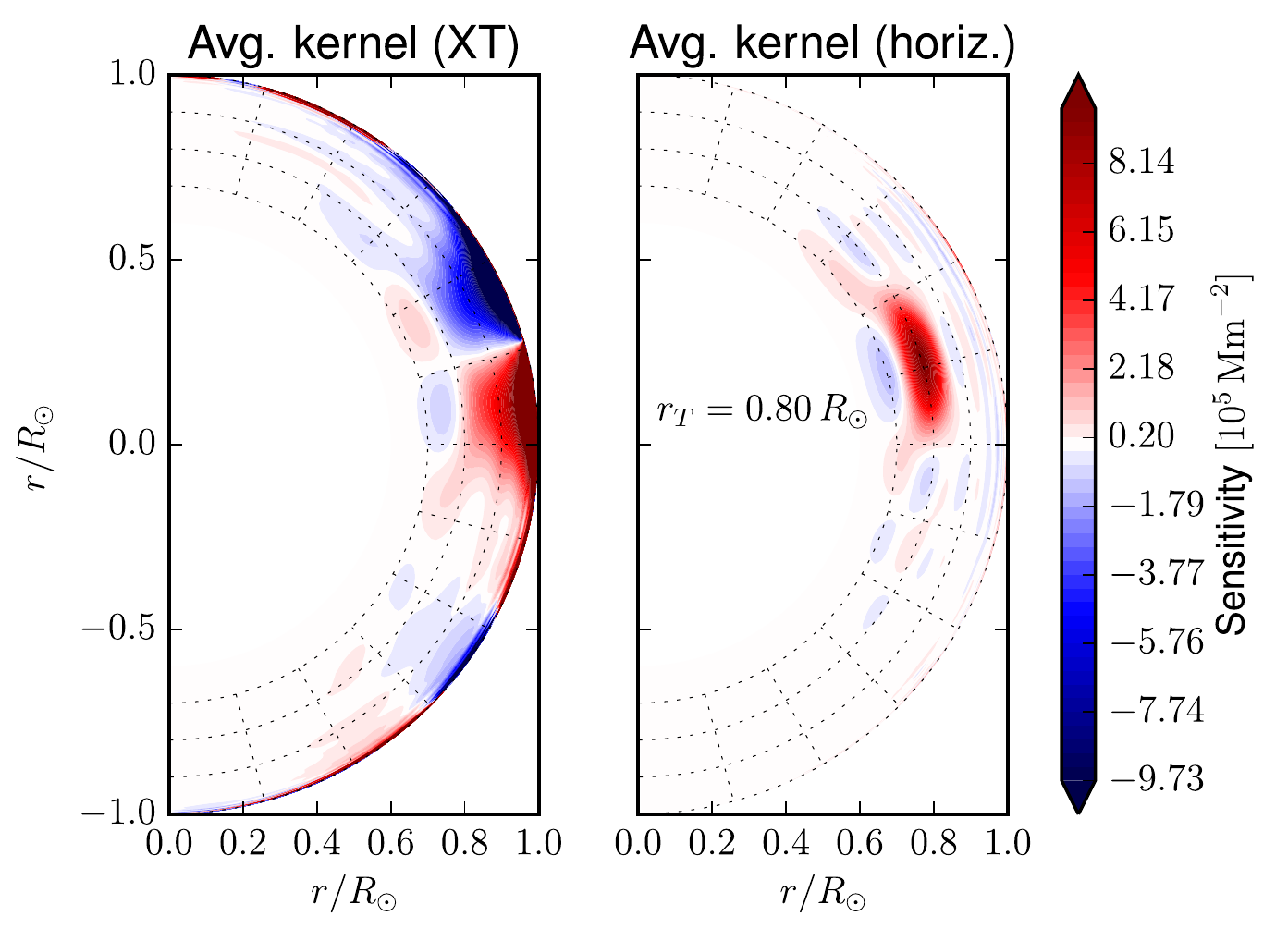}%

\includegraphics[width=0.5\textwidth]{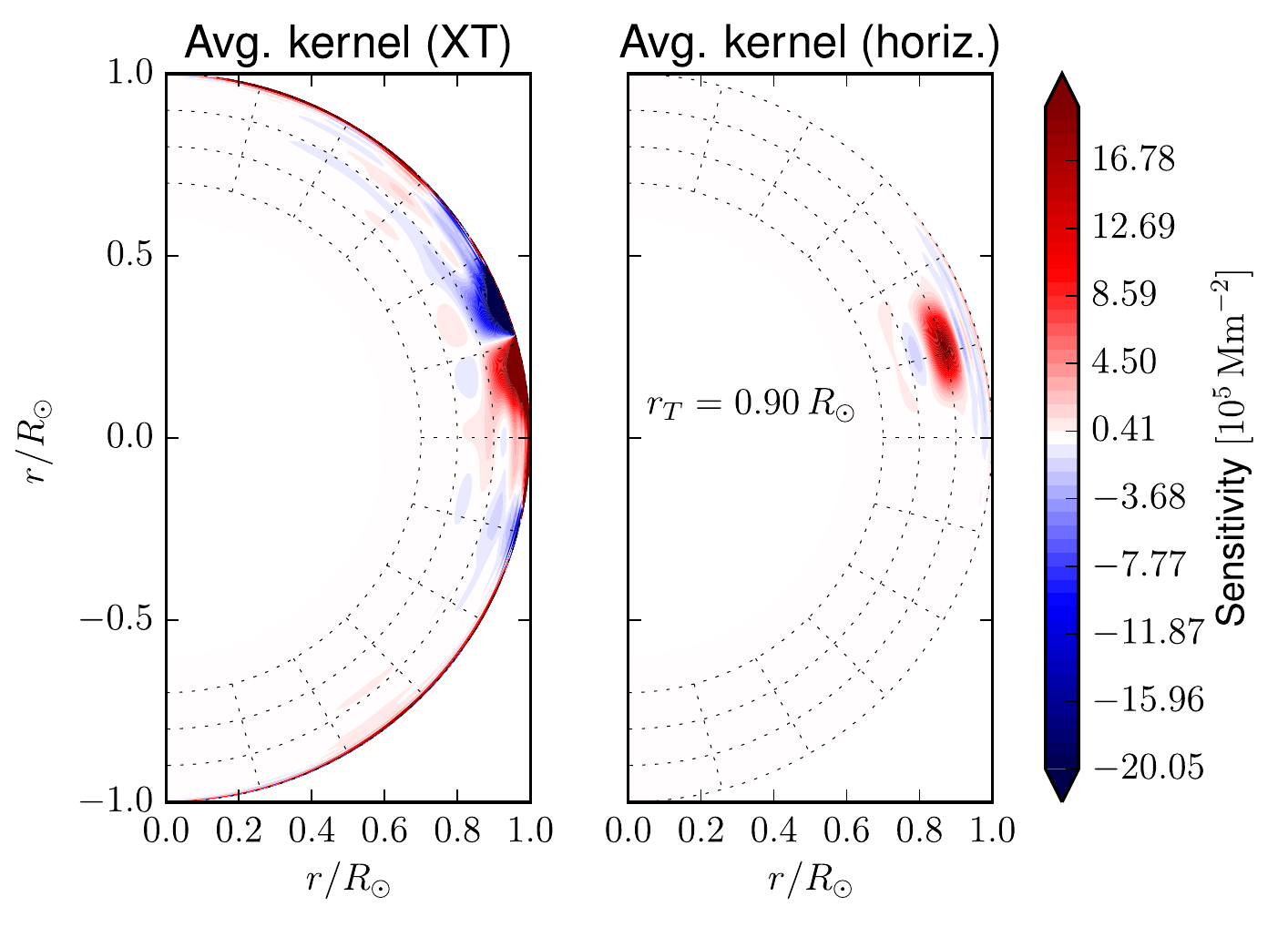}%
\includegraphics[width=0.5\textwidth]{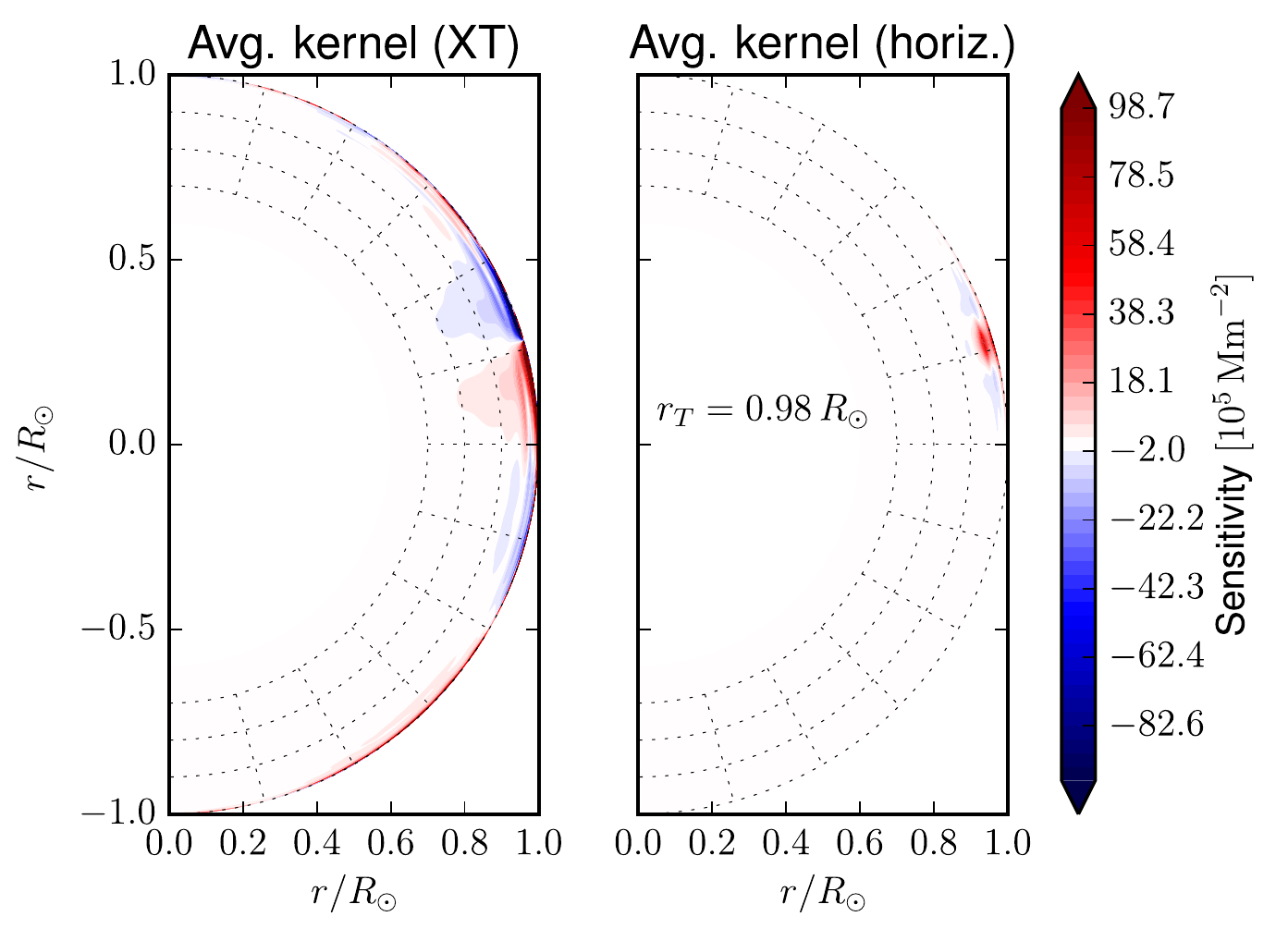}%

\caption{Averaging kernels for the inversion result presented in Figure~\ref{figsolabornall} at a target latitude of $16.3\degr$ and target depths of $r_T = 0.7,\,0.8,\,0.9,\,0.98\,R_\sun$ (from top left to bottom right). In each panel, the left subpanel shows the cross-talk averaging kernel $\cK_r$ and the right subpanel shows the averaging kernel $\cK_\theta$. Note that the cross-talk kernels are highly saturated for visibility of the horizontal averaging kernels, with maximum values of the cross-talk averaging kernels being 120, 100, 70, and 14 times larger (from top left to bottom right).}%
\label{figxtavgKs}%
\end{figure*}

Furthermore, although the cross-talk was not considered in the inversion procedure, it is possible to compute cross-talk averaging kernels, i.e., kernels for the influence of the radial flow on the inversion for horizontal flow,
\begin{align}
\cK_r(r,\theta;r_T,\theta_T) = \sum_i K_r^i(r,\theta)w_i(r_T,\theta_T).
\end{align}
For a series of target depths, we show the averaging kernel for the horizontal flow (right subpanel) and the cross-talk kernel for the radial into the horizontal component (left subpanel) in Figure~\ref{figxtavgKs}. We note that for each target depth plotted, both kernels are shown using the maximum values of the averaging kernel as a scale, with the maximum values of the cross-talk averaging kernels being 120, 100, 70, and 14 times larger for the given target depths $r_T$ of 0.7, 0.8, 0.9, and 0.98 $R_\sun$. These values are much larger than the ones obtained by \cite{Svanda2011} in an inversion for subsurface flows in Cartesian geometry.

\begin{figure}%
\includegraphics[width=\columnwidth]{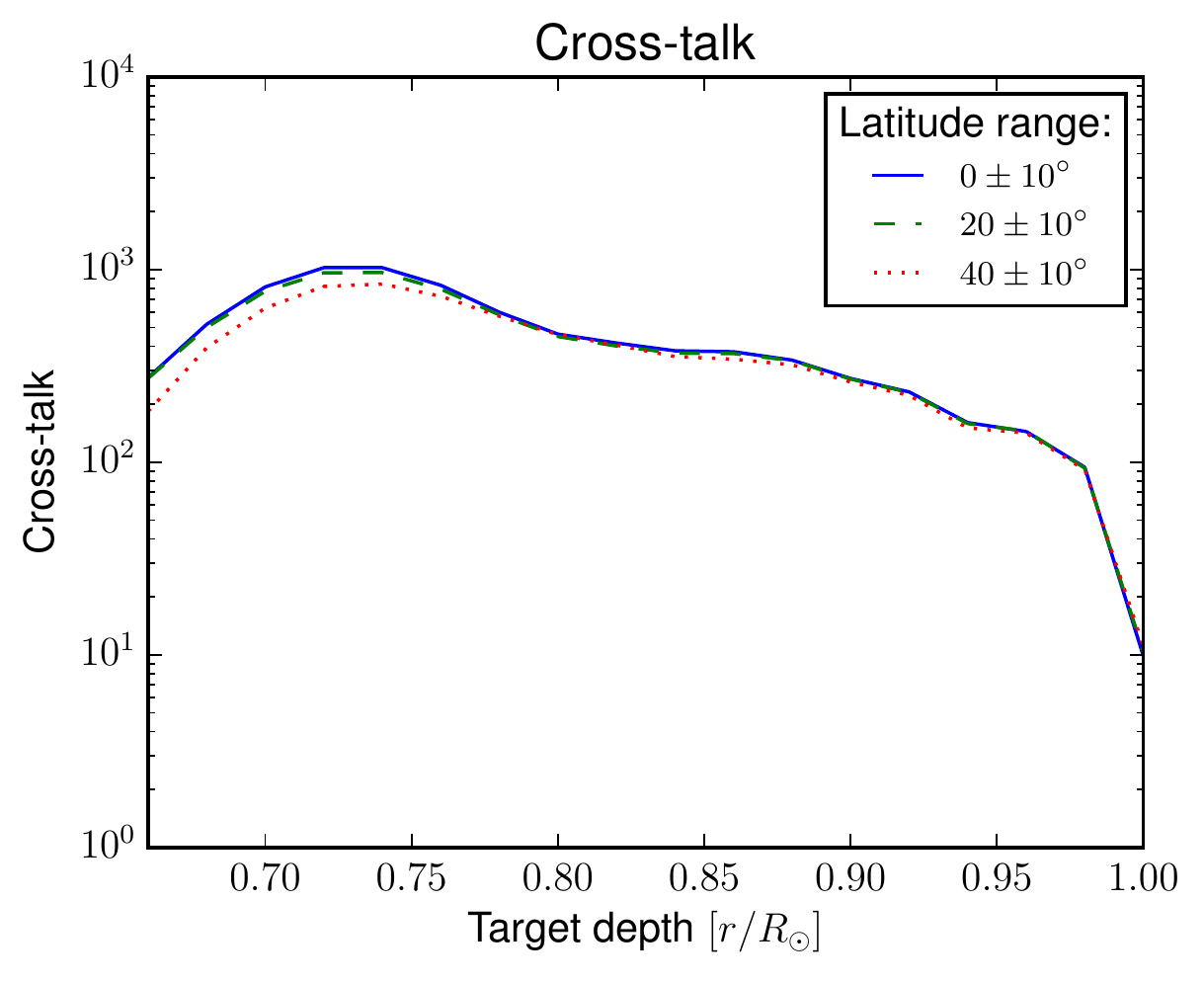}

\caption{Normalized cross-talk for the inversion result presented in Figure~\ref{figsolabornall}. See text for the definition of the cross-talk and a discussion of its magnitude.}%
\label{figxt1d}%
\end{figure}

\begin{figure}%
\includegraphics[width=\columnwidth]{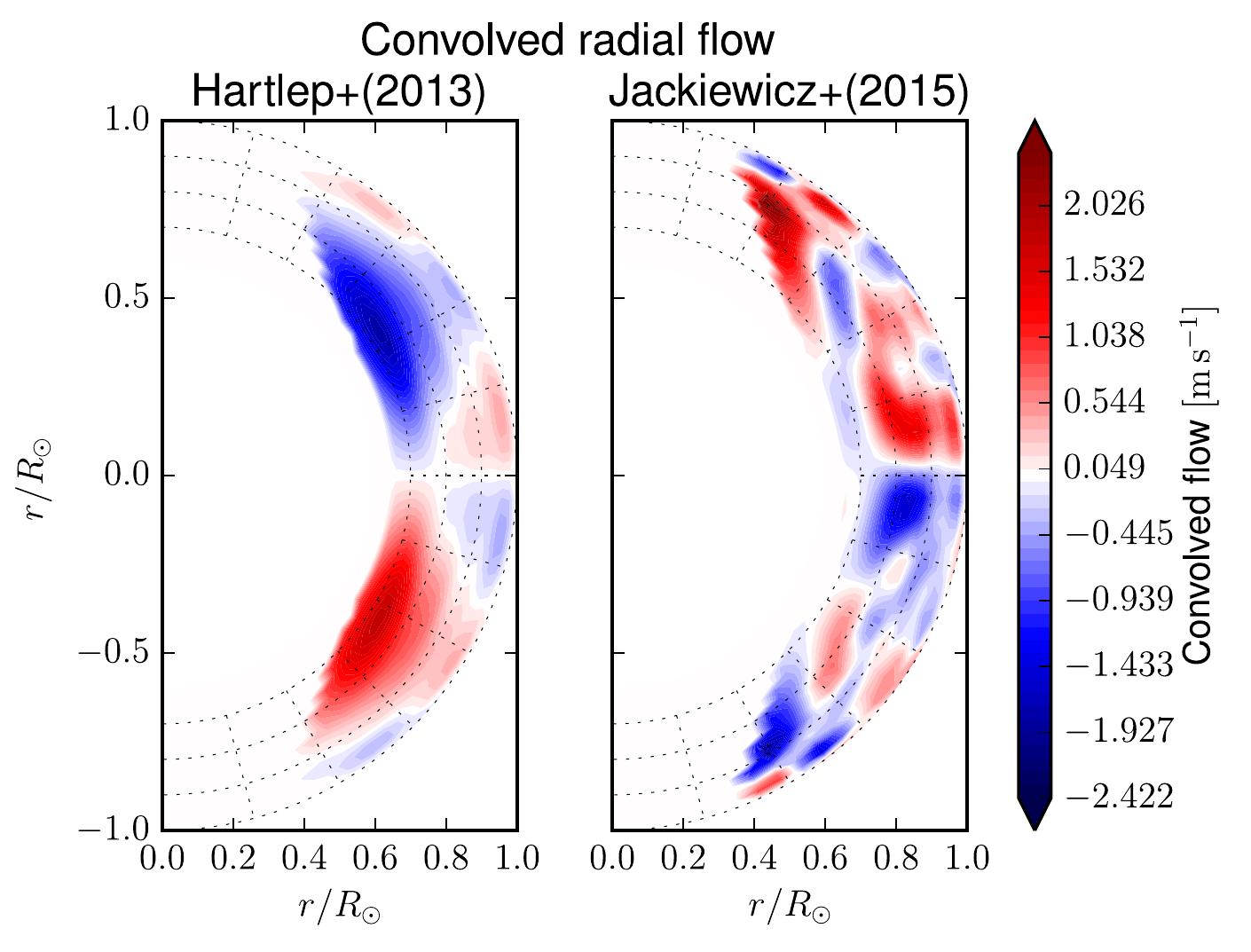}

\caption{Convolution of two example radial flow profiles with the cross-talk averaging kernels $\cK_r$. The results shown were obtained using the radial flow profile from \citet[][left, divided by a factor of 36 to obtain solar values]{Hartlep2013} and using the radial flow profile obtained from the horizontal inversion result from \citet[][see MF4, third column in their Figure~5]{Jackiewicz2015} using a mass-conservation constraint. The convolved flows displayed here are estimates for the impact of possible radial flow profiles on the inversion for horizontal flow.}%
\label{figavgKrconv}%
\end{figure}

We therefore study the nature and impact of the cross-talk in more detail, see Figure~\ref{figxt1d}, where the normalized cross-talk,
\begin{align}
\XT_{\rm{norm}}  = \frac{\iint  \cK_r^2 \; r \sin(\theta)\,\id\theta \,\id r}{\iint  T ^2 \; r \sin(\theta)\,\id\theta \,\id r}
\end{align}
is displayed as a function of depth, for a number of latitude bins. For example, a normalized cross-talk of 100 means that radial flows may contribute to the inversion for horizontal flow even if they are about $\sqrt{100}=10$ times smaller than the actual horizontal flow. In principle, even a small magnitude radial flow may thus leak into the inversion for the horizontal flow in such a case. In order to answer the question whether such values for the cross-talk introduce a large contribution of the radial flow to the inverted horizontal flow, we convolved the cross-talk averaging kernels with two choices of exemplary radial flow fields $v_m^{\rm{orig}}\, (m=r)$,
\begin{align}
v_\theta&^{\rm{conv}(m)}(r_T,\theta_T)  \nonumber \\
& =\iint  \cK_m(r,\theta;r_T,\theta_T) \, v_m^{\rm{orig}}(r,\theta) \; r \sin\theta\,\id\theta \,\id r. \label{eqconvflow}
\end{align}
The resulting contribution of the radial flow signal to the inversion for $v_\theta$ is displayed in Figure~\ref{figavgKrconv}. For the left panel, we chose $v_r^{\rm{orig}}$ to be the radial flow component of the single-cell meridional flow profile employed in the simulation of \cite{Hartlep2013}, divided by a factor of 36 in order to mimic a realistic magnitude of a solar-type flow. For the right panel, we chose the radial flow component that was obtained by \cite[][see MF4, third column in their Figure 5]{Jackiewicz2015} from the inverted horizontal flow by applying mass conservation. In both cases, the contribution of the radial flow component to the inversion for $v_\theta$ has maximal values of around $2\,\mpersec$, with typical values of about $1\,-\,1.5\,\mpersec$. We may thus expect small contributions from the radial flow component, but it is likely that they do not alter the inversion result at a large scale. However, one should be aware of this contribution.

\section{Horizontal SOLA Inversion including Full Covariance and Cross-talk}
\label{secfullinv}

Refined inversions are now performed by including the full covariance matrix and a regularization term for cross-talk into the inversion problem. A similar problem has been studied for inversions for 3D  flows near the surface in Cartesian geometry by, e.g., \cite{Svanda2011,Svanda2013Validation} and \cite{DeGrave2014QS}.

\subsection{The Full Inverse Problem}

In the full SOLA inverse problem, we invert for a flow component $k\in\{r,\theta\}$,
\begin{equation}
v_k^{\rm{inv}}(r_T,\theta_T) = \sum_i w_i^k(r_T,\theta_T) \, \delta\tau_i,
\end{equation}
by trying to match the averaging kernels,
\begin{align}
\cK_m^k(r,\theta;r_T,\theta_T) = \sum_i K_m^i(r,\theta)w_i^k(r_T,\theta_T),
\end{align}
to a target kernel $T_k^m = \delta_{m,k} T$, where $T$ is the same target function from above. When inverting for a flow component $k$, we thus also intend to match the averaging kernel of the opposite component $m\neq k$ to zero. Therefore, a regularization for the cross-talk of the flow component $m\neq k $ into $k$,
\begin{align}
\XT_k = \iint \left(\cK_{m\neq k}^k\right)^2  \; r \sin(\theta)\,\id\theta \,\id r,
\end{align}
is added to the inversion problem outlined in Section~\ref{secdatainv} (see, e.g, \cite{Svanda2011} for the equivalent formulation in Cartesian geometry). The cost function thus becomes
\begin{align}
\chi_k(w_i^k;\mu) &= \MF + \mu \,\err^2 + \nu \,\XT_k 
\end{align}
which is to be minimized subject to the constraints
\begin{align}
\iint \cK_m^k \; r \sin(\theta)\,\id\theta \,\id r = \delta_{km}. \label{eqnormavgKs}
\end{align}

As in the inversion without cross-talk, inversion parameters are chosen in a preparatory inversion using a coarser spatial grid for the target locations. The FWHMs of the target functions are the same.

In the following, we invert for the horizontal component ($k=\theta$) and show the inversion results for parameters obtained using different strategies. In all cases, we first set an upper limit to the misfit of 0.2, which is the same as in the inversion without cross-talk in the previous section.

\begin{figure*}%
\includegraphics[height=0.26\textwidth]{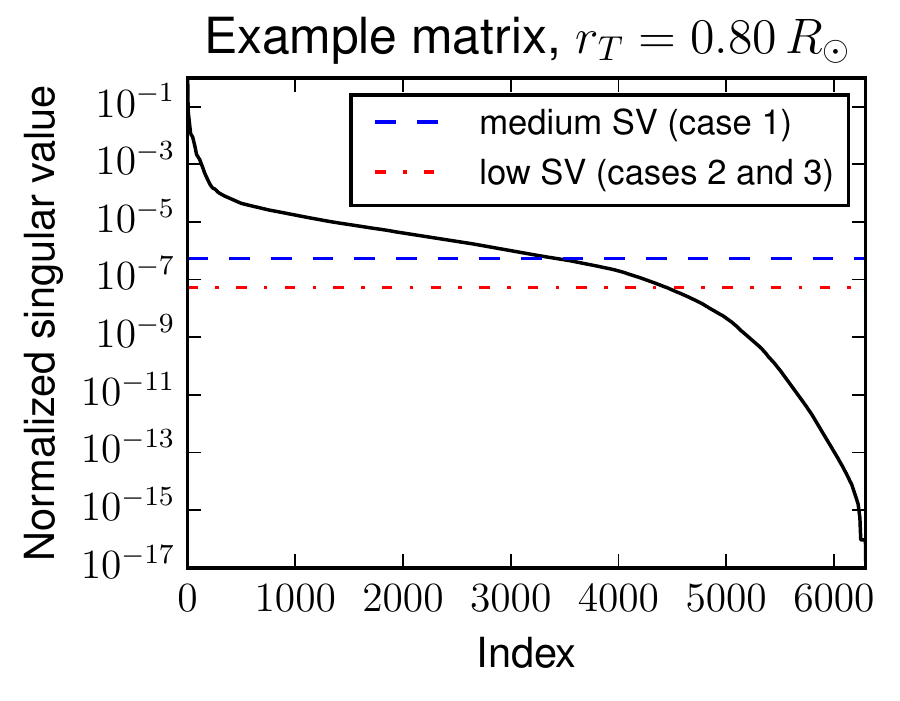}%
\includegraphics[height=0.26\textwidth]{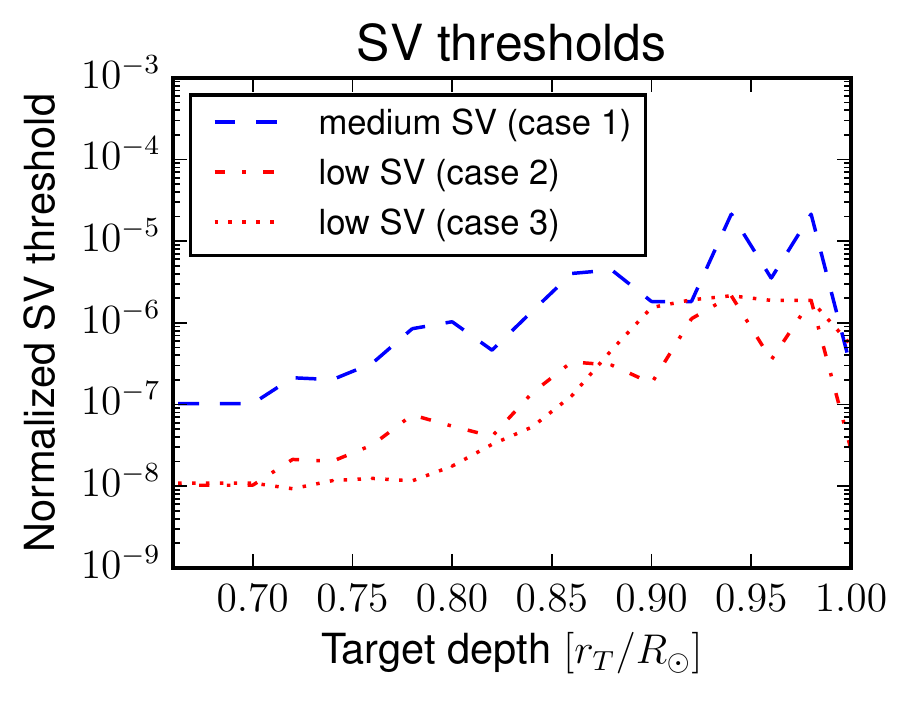}%
\includegraphics[height=0.26\textwidth]{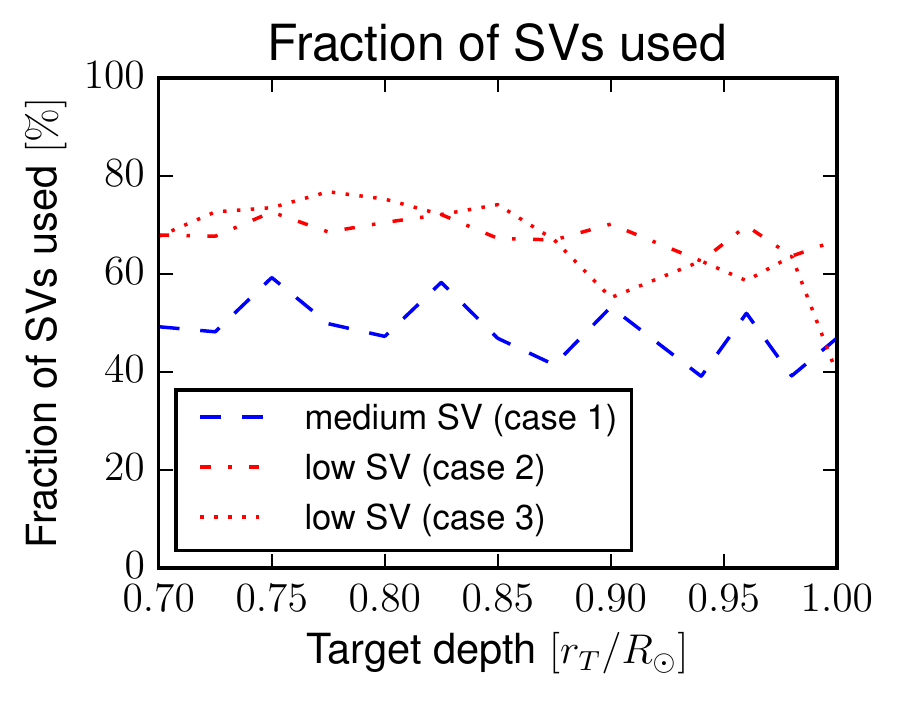}%
\caption{Singular values (SVs) and thresholds used in the singular value decomposition of the refined inversions in Section~\ref{secfullinv}. Shown are SVs for a typical matrix at a target radius of $r_T=0.8\,R_\sun$ (left panel), the dependence of thresholds on target depth (middle panel), and the fraction of SVs used in the matrix inversion (right panel). Thresholds used for cases 2 and 3 are not identical in the middle and right panels because a different selection of inversion parameters leads to different matrices being inverted.}%
\label{fig_sv}%
\end{figure*}

As the inversion results presented here are found to depend on the threshold used in the singular value decomposition (SVD), we show the SVs of a matrix used in the inversion at an example target depth of $r_T=0.8\,R_\sun$ in the left panel of Figure~\ref{fig_sv}. As the distribution of SVs depends on target depth, the thresholds at each target depth are chosen relative to the point of highest negative curvature, which is found by fitting a high-order polynomial to the curve. Therefore, the thresholds and the fraction of SVs used in the matrix inversion also depend on target depth; see middle and right panel of Figure~\ref{fig_sv}.

\subsection{Results: Controlling Misfit and Errors}
\label{secfullinv_maxErr}

In addition to the misfit, we aim to keep the errors under control in the first inversion. After choosing a threshold for the SVs, we also set a maximum threshold of $1.0\,\rm{m}\,\rm{s}^{-1}$ for the inversion errors as in Section~\ref{secfirstcomp}. We then search the parameter space for the best cross-talk available. 

\begin{figure*}%

\begin{center}

\includegraphics[width=\textwidth]{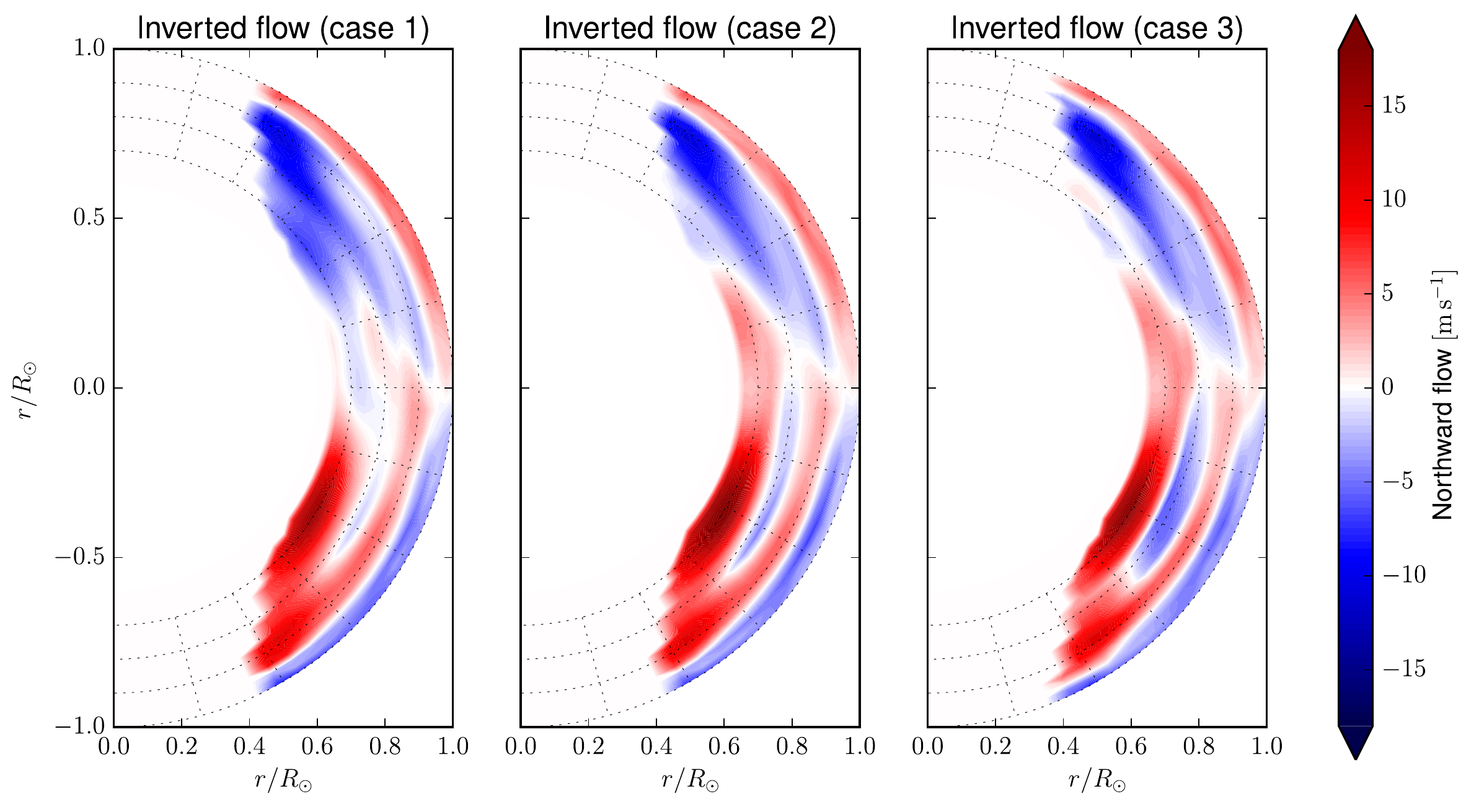}
\end{center}

\caption{Main result: inverted meridional flow profiles from the refined SOLA inversion using the full covariance matrix and a regularization term for cross-talk. For the different cases, we used either a medium threshold (case 1) or a low threshold (cases 2 and 3) in the SVD, while trade-off parameters were searched for with a threshold for maximum error (cases 1 and 2) or maximum cross-talk (case 3).}%
\label{fig_comp_flows}%
\end{figure*}

As a first case, we consider an SV threshold chosen to be just above the end of the rather flat plateau of SVs in the left panel of Figure~\ref{fig_sv} (indicated in Figure~\ref{fig_sv} by ``medium SV''). Here, about half of the SVs are used in the matrix inversion.
In this case, we obtain an inverted flow profile that is very similar to the one obtained by \cite{Jackiewicz2015}; see the left panel in Figure~\ref{fig_comp_flows}, subsequently termed ``case 1''. Compared to the initial result shown in Figure~\ref{figsolabornall}, however, it shows less fluctuations as a function of latitude in each hemisphere. This difference is found to be due to the use of the full covariance matrix.

In a second case, we choose an SV threshold smaller by a factor of 10 just at the edge of the plateau of SVs in the left panel of Figure~\ref{fig_sv} indicated by ``low SV''. Here, about 70 \% of SVs are used. In this case, the inverted flow profile shows some noteworthy features; see case 2 in Figure~\ref{fig_comp_flows}. 
Most notably, there is an additional extended poleward flow branch visible at about $0.8\,R_\sun$ in the southern hemisphere. This flow component gives rise to a multi-cell structure of the flow stacked radially. Depending on the associated radial flow, the inverted meridional flow corresponds to a double-cell profile \citep[as in, e.g., Fig. 2 in ][]{Hazra2014} or to three flow cells stacked radially \citep[as in, e.g., Fig. 4 in ][]{Hazra2014}. At the same time, no additional flow cells are visible in the northern hemisphere. At the central location of the additional poleward flow structure in the southern hemisphere at about $0.8\,R_\sun$ and $30\degr$ latitude, a patch with equatorward flow is also visible in the result for case 1, in the initial result in Figure~\ref{figsolabornall}, and in the result obtained by \cite{Jackiewicz2015} using ray kernels.

\begin{figure*}%

\includegraphics[width=0.33\textwidth]{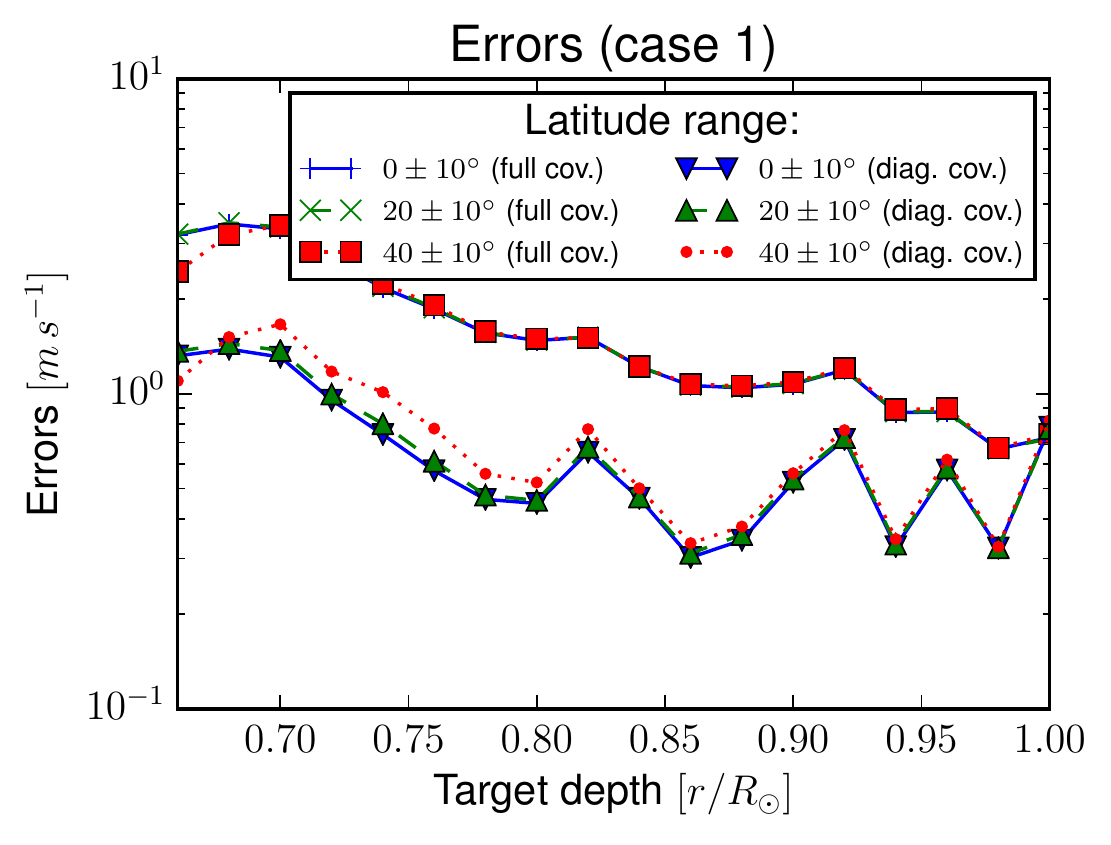}%
\includegraphics[width=0.33\textwidth]{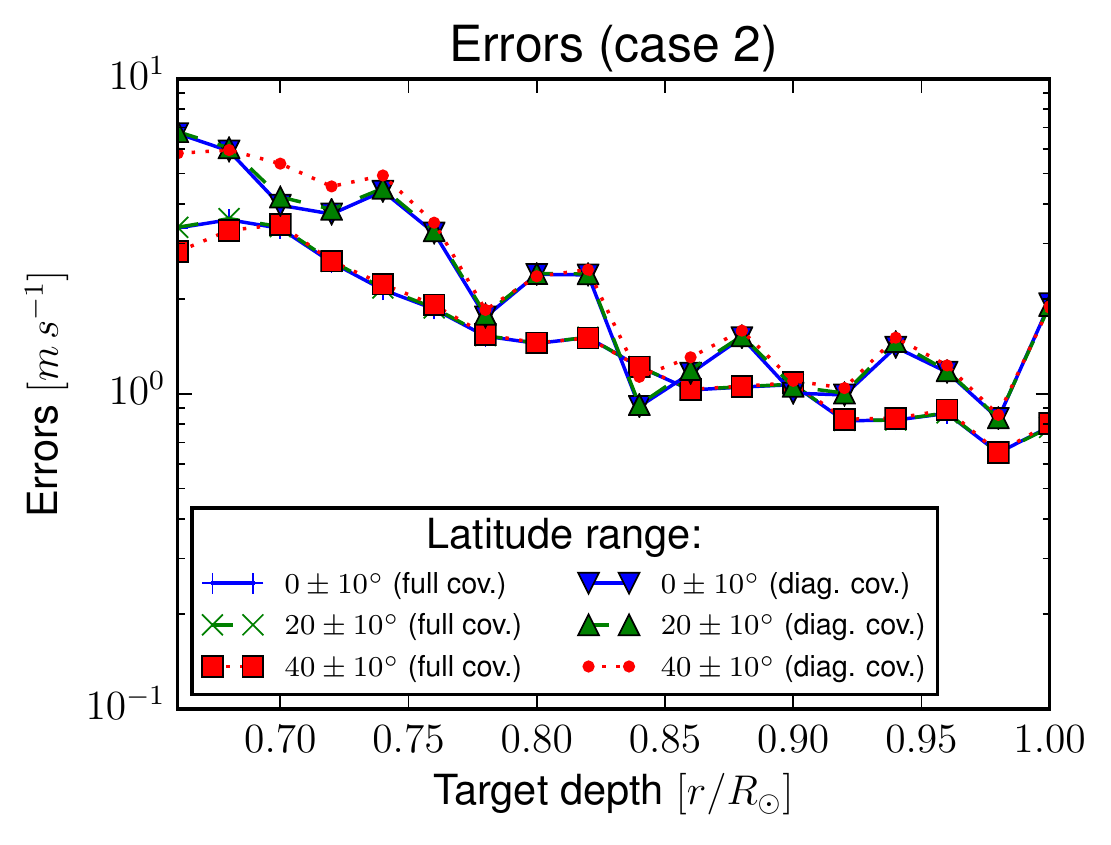}%
\includegraphics[width=0.33\textwidth]{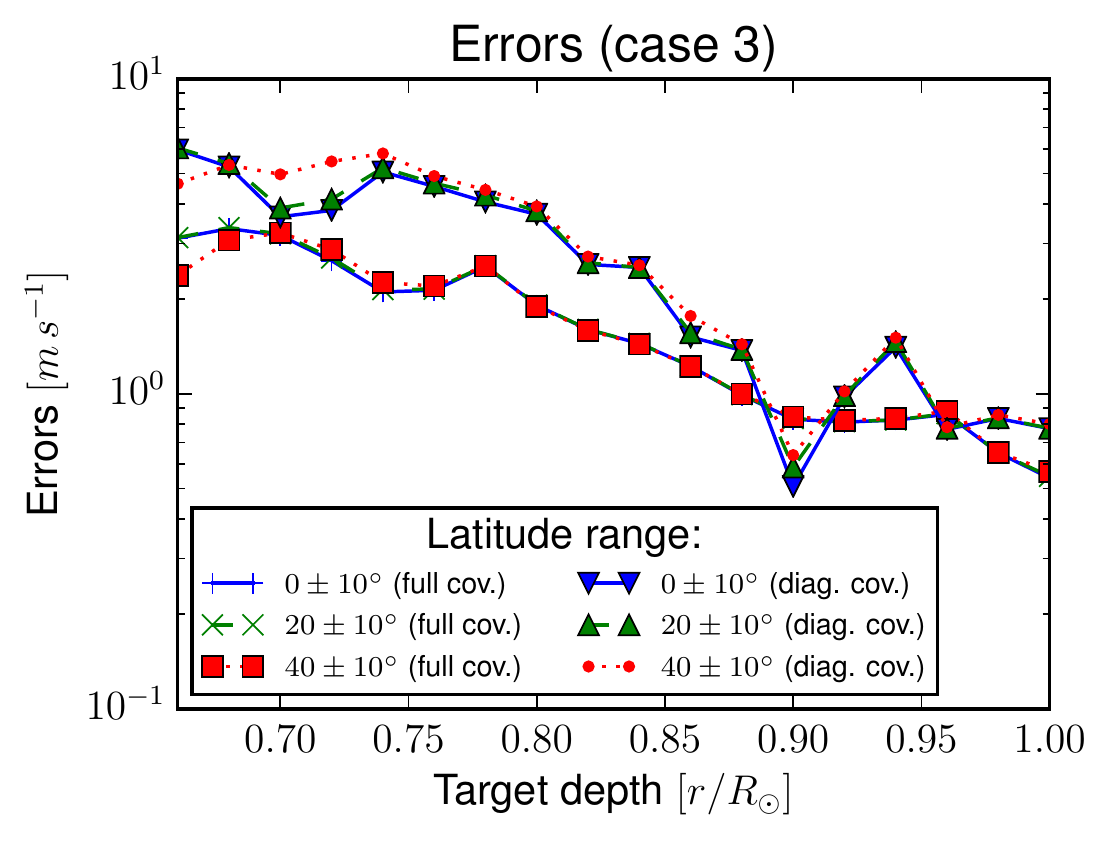}%

\caption{Errors for the inverted flows shown in Figure~\ref{fig_comp_flows}. Errors were propagated through the inversion using the assumption of uncorrelated travel-time measurements (diagonal covariance) and taking the full covariance of the measurements into account; see legends. For a lower threshold in the SVD (cases 2 and 3), the errors from the full covariance become smaller than the errors from the diagonal one.}%
\label{fig_comp_errors}%
\end{figure*}

Inversion errors using the full and a diagonal covariance matrix are shown in Figure~\ref{fig_comp_errors}. With the full covariance matrix, we now obtain errors that are generally smaller than when using a diagonal covariance matrix by about a factor of 1.5 in case 2, with a few exceptions. In the case of the medium SV threshold (case 1), however, the errors from the full covariance matrix are on average about twice as large as the ones from the diagonal covariance matrix, similar to the initial inversion result (see Figure~\ref{figerrscatter}). Note that the errors from the diagonal covariance matrix increase on average by about 35\% if the diagonal values of the covariance matrices are assumed to be identical after rebinning in latitude as in Section~\ref{secfirstcomp}.

Furthermore, one can see that the cross-talk decreased (see Figure~\ref{fig_comp_xt}, case 1 and 2), by about a factor of 3 especially at larger target depths compared to the cross-talk for the initial result; see Figure~\ref{figxt1d}. As a consequence, the convolution of the two examples of radial flow profiles discussed in Section~\ref{secfirstcomp} with the radial averaging kernels decreases by over 30\% to below $1.4\,\mpersec$.

Concerning the cross-talk, it thus seems that the horizontal inversion without regularization for cross-talk already achieved results that are near optimum, or in other words, that the cross-talk is not a major concern for the horizontal inversion. A similar conclusion was reached by \cite{Svanda2011} in the case of a Cartesian inversion for the horizontal component of a 3D flow.

For misfit and errors, the distribution of values over the whole target grid is similar to that presented in Figure~\ref{figsolabornall} for all results presented in this section.

\subsection{Results: Controlling Misfit and Cross-talk}
\label{secfullinv_maxXT}

Still imposing the condition of a maximum misfit of 0.2 and the low SV threshold (same as in case 2 for a given matrix), we now take a slightly different strategy for choosing the trade-off parameters. Motivated by the analysis shown in Figure~\ref{figavgKrconv}, we first choose a maximum value of 100 for the cross-talk (case 3 in the following). This means that the cross-talk averaging kernel $\cK_r^\theta$ has a mean magnitude of about 10 times larger than the averaging kernel $\cK_\theta^\theta$. If this condition cannot be met, both maximum misfit and cross-talk are increased step by step until a condition is found that can be met in the parameter space. We finally search for the regularization parameter that gives minimal errors.

The inversion results for this case are also displayed in Figure~\ref{fig_comp_flows}. In the whole region considered, the inverted flow is very similar to the one obtained in case 2 above. It is noteworthy, however, that the multi-cell structure in the southern hemisphere is even more pronounced in case 3 compared to case 2.

\begin{figure*}

\includegraphics[width=0.33\textwidth]{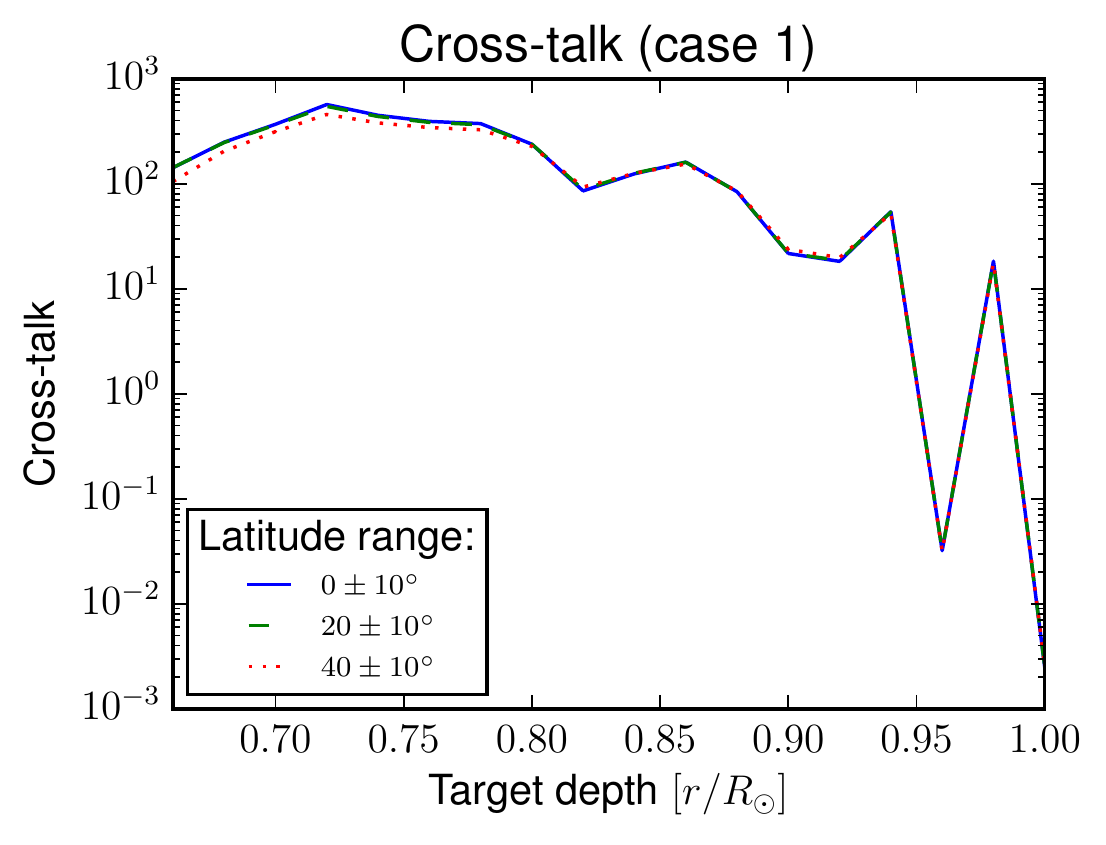}%
\includegraphics[width=0.33\textwidth]{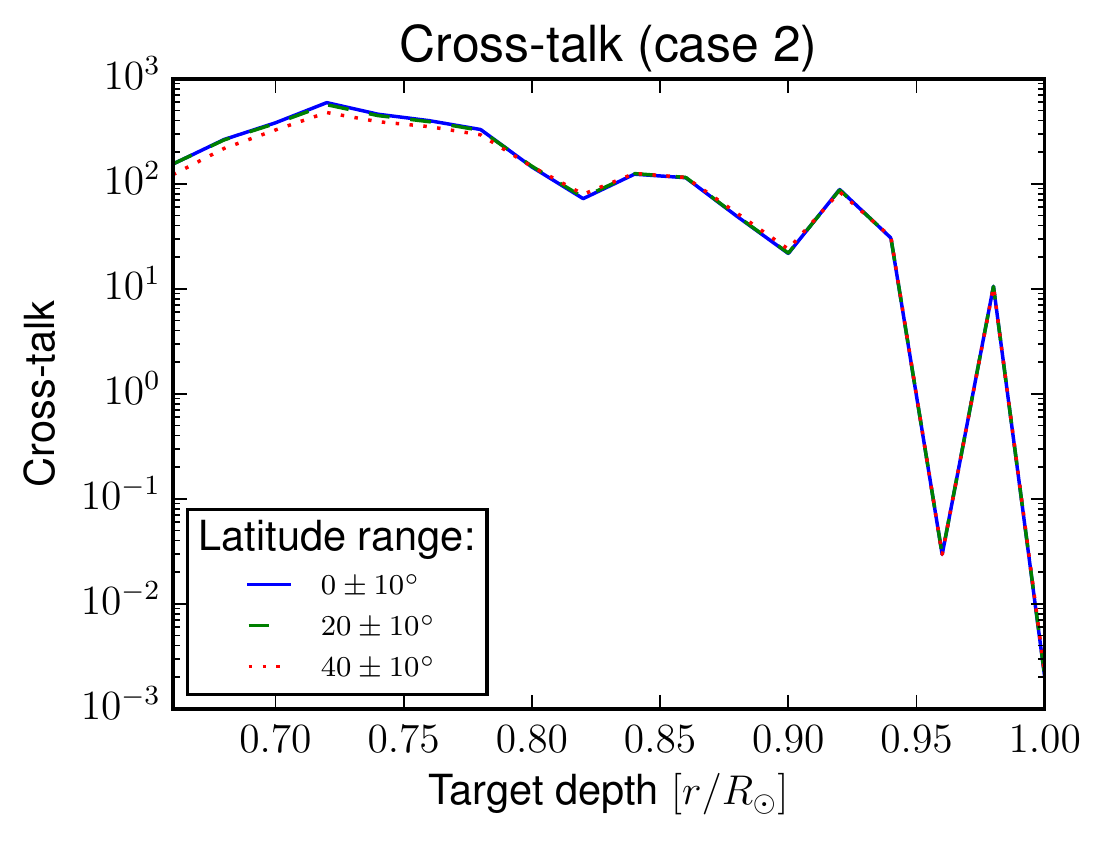}%
\includegraphics[width=0.33\textwidth]{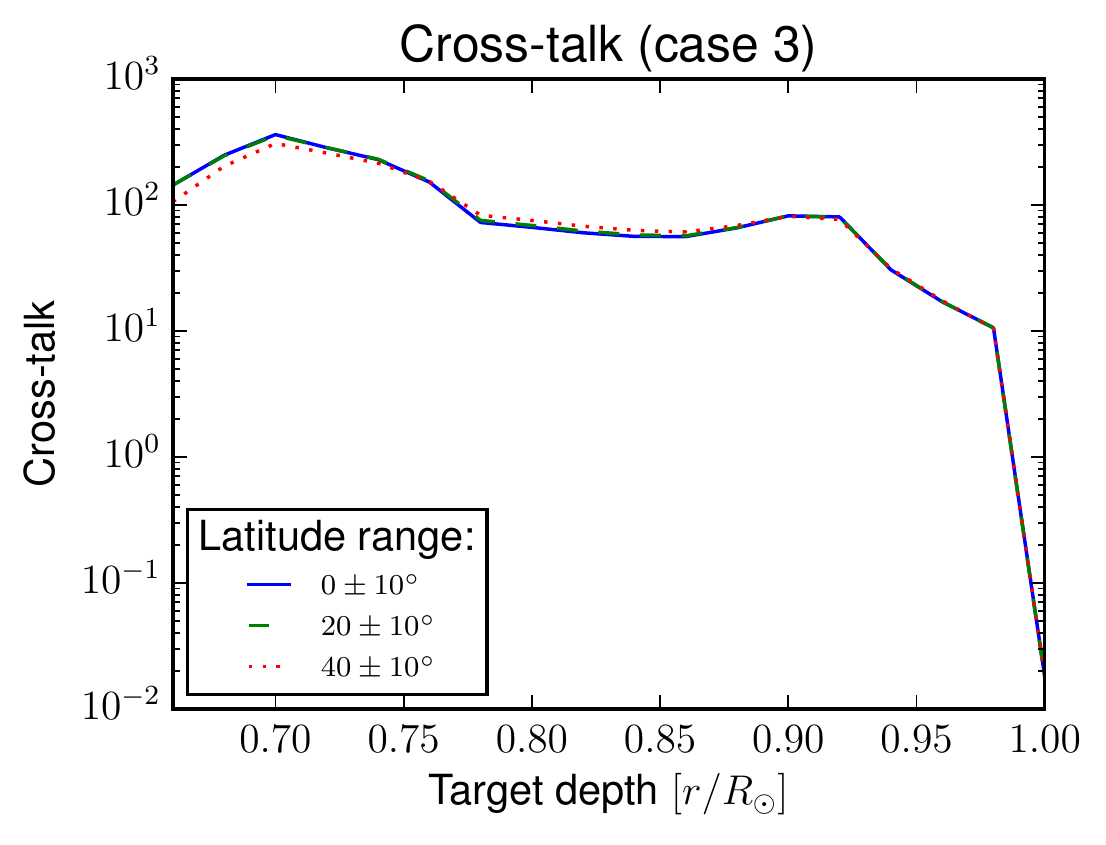}%

\caption{Normalized cross-talk for the inversion results presented in Figure~\ref{fig_comp_flows}.}%
\label{fig_comp_xt}%
\end{figure*}

\begin{figure*}%

\includegraphics[width=0.5\textwidth]{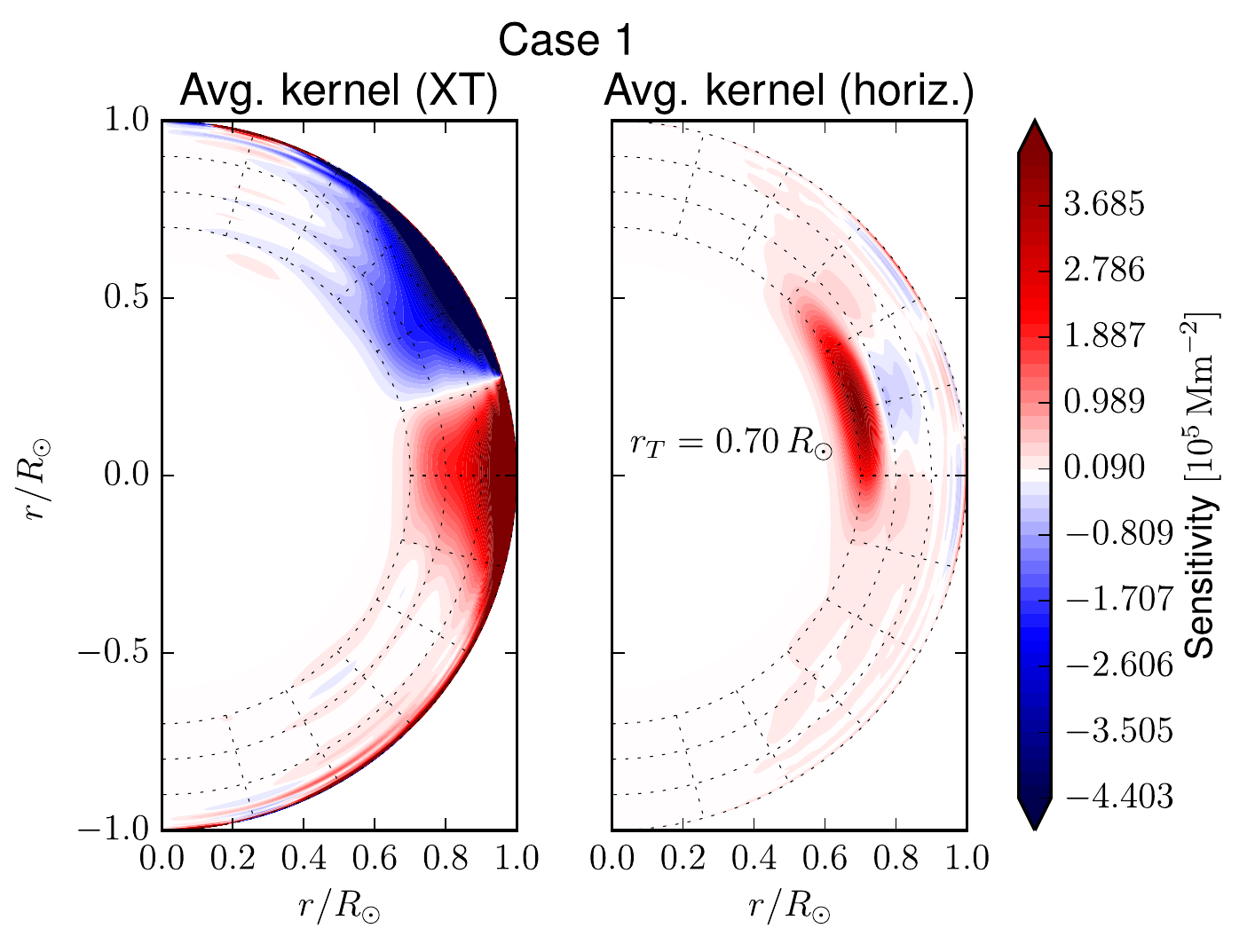}%
\includegraphics[width=0.5\textwidth]{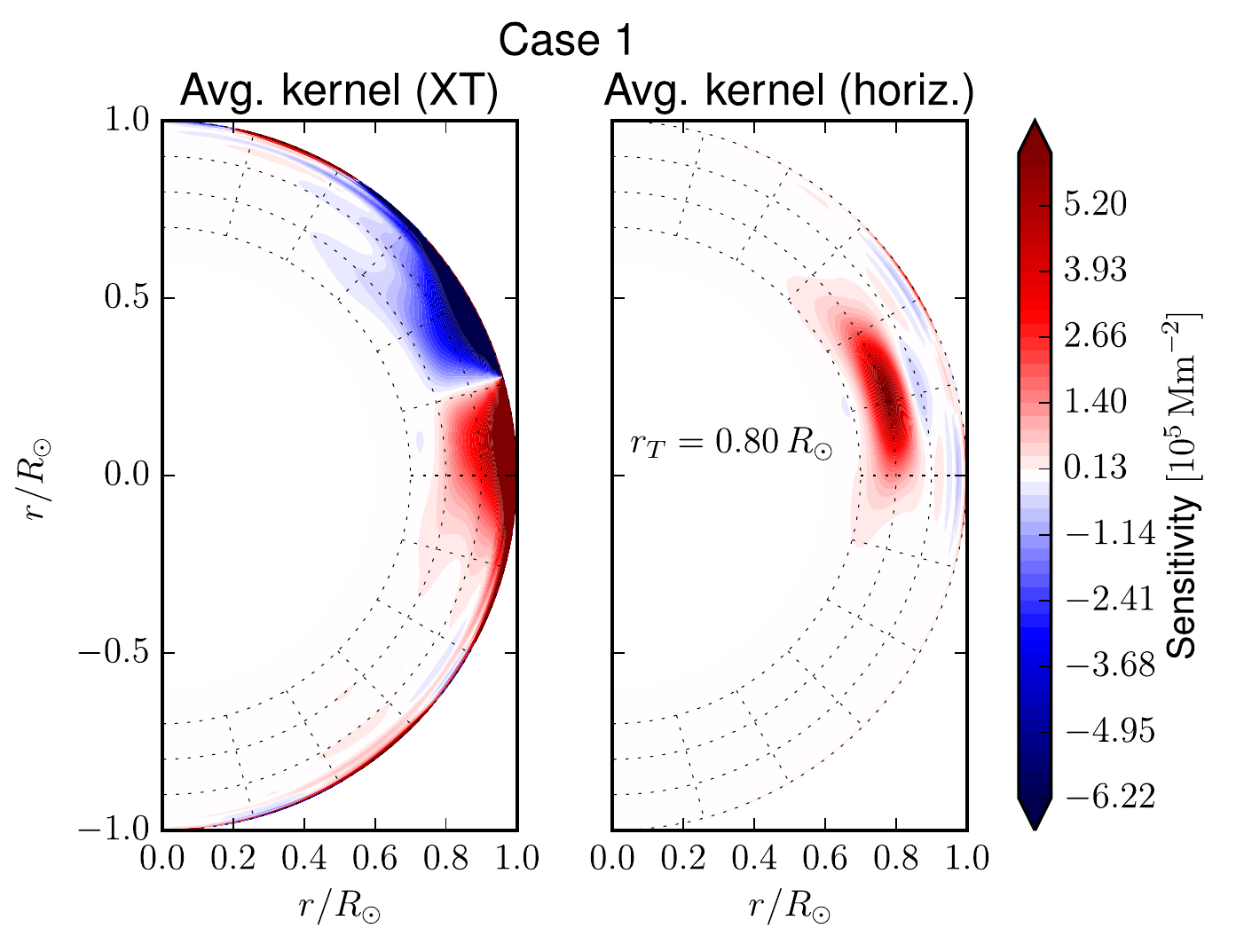}%

\includegraphics[width=0.5\textwidth]{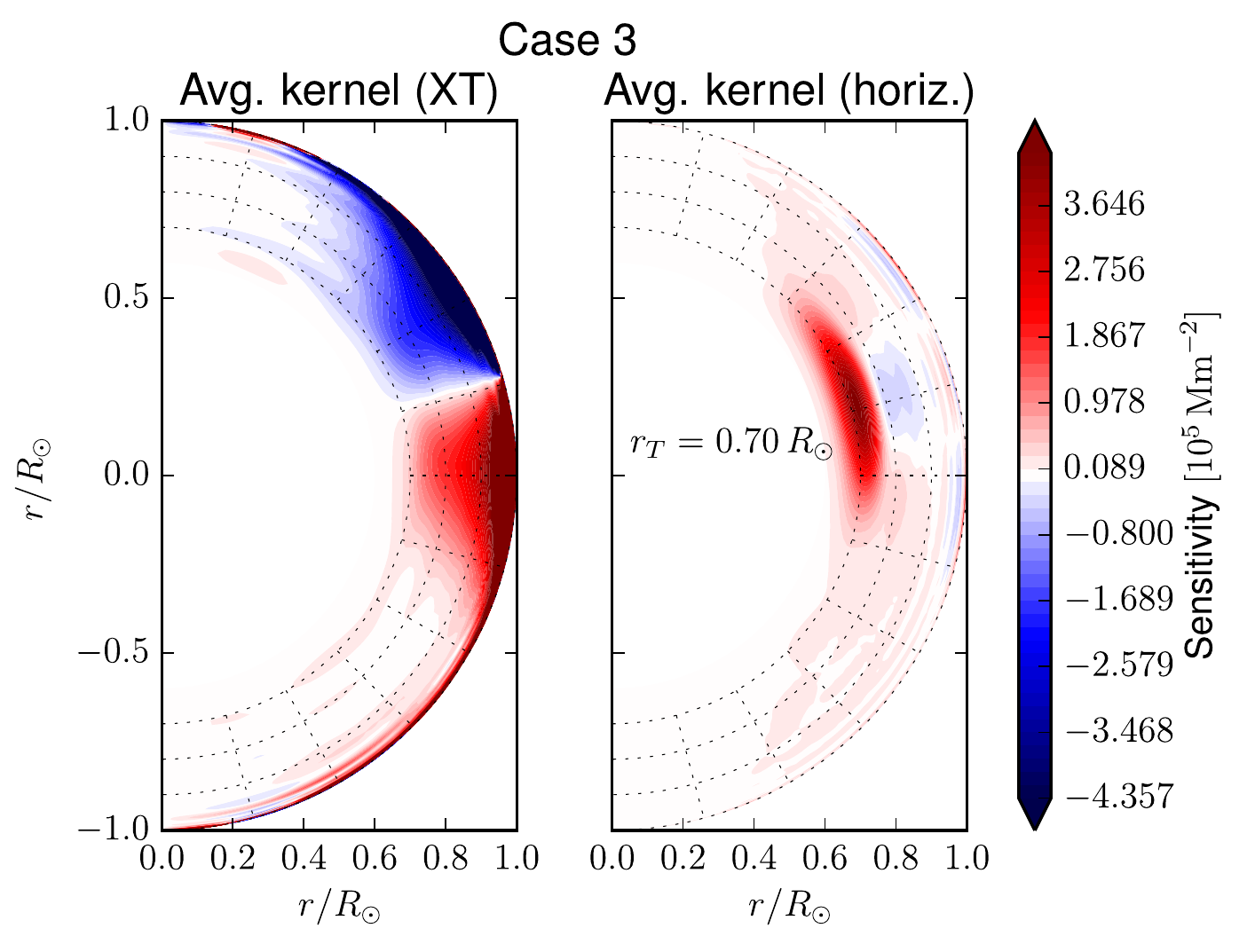}%
\includegraphics[width=0.5\textwidth]{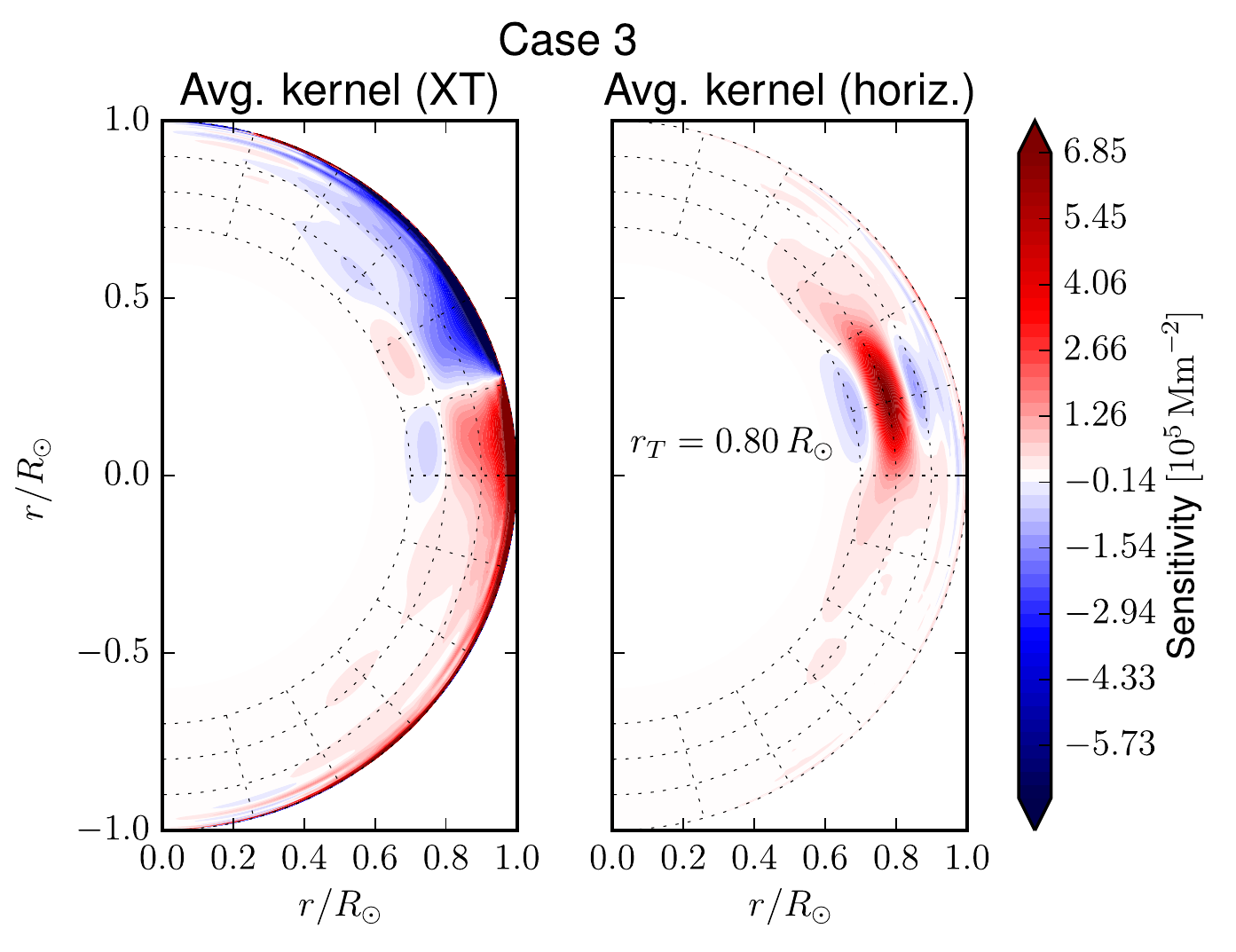}%
	
\caption{Averaging kernels for the inversion result presented in Figure~\ref{fig_comp_flows} (case 1: top row, case 3: bottom row), a target latitude of $16.3\degr$ and target depths of $r_T = 0.7\,R_\sun$ (left column) and $r_T=0.8\,R_\sun$ (right column). In each panel, the left subpanel shows the cross-talk averaging kernel $\cK_r^\theta$ and the right subpanel shows the averaging kernel $\cK_\theta^\theta$. Note that the cross-talk kernels are highly saturated for visibility of the horizontal averaging kernels.}%
\label{figxtavgKsBest}%
\end{figure*}

This result was obtained although the dependence of the errors and the cross-talk as a function of depth changed only slightly from case 2 to case 3; see Figures~\ref{fig_comp_errors} and \ref{fig_comp_xt}. 
In the near-surface regions, we are able to obtain slightly smaller error bars that increase with depth, peaking at about $3\,\mpersec$ at the bottom of the convection zone. The cross-talk, on the other hand, decreased slightly in regions below $0.8\,R_\sun$, and it increased slightly around $0.9\,R_\sun$. As a result, the magnitude of the cross-talk averaging kernels changed accordingly, which is visible in the right panels of Figure~\ref{figxtavgKsBest} with a smaller saturated region at $r_T=0.9\,R_\odot$ in case~3 compared to case~1.

We note that to obtain the third result presented here, we allowed the misfit to increase near the bottom of the convection zone by about 20\%. We observed that only a slight decrease of the misfit would have increased the errors or the cross-talk by a large amount there. This increase in the misfit is not expected to largely alter the quality of the inversion result compared with the one obtained in Section~\ref{secfullinv_maxErr}, as this change is barely visible in the averaging kernels (horizontal kernels in the left panels of Figure~\ref{figxtavgKsBest}).

\section{Discussion}
\label{secdiscussion}

\begin{figure*}%
\begin{center}

\includegraphics[width=\textwidth]{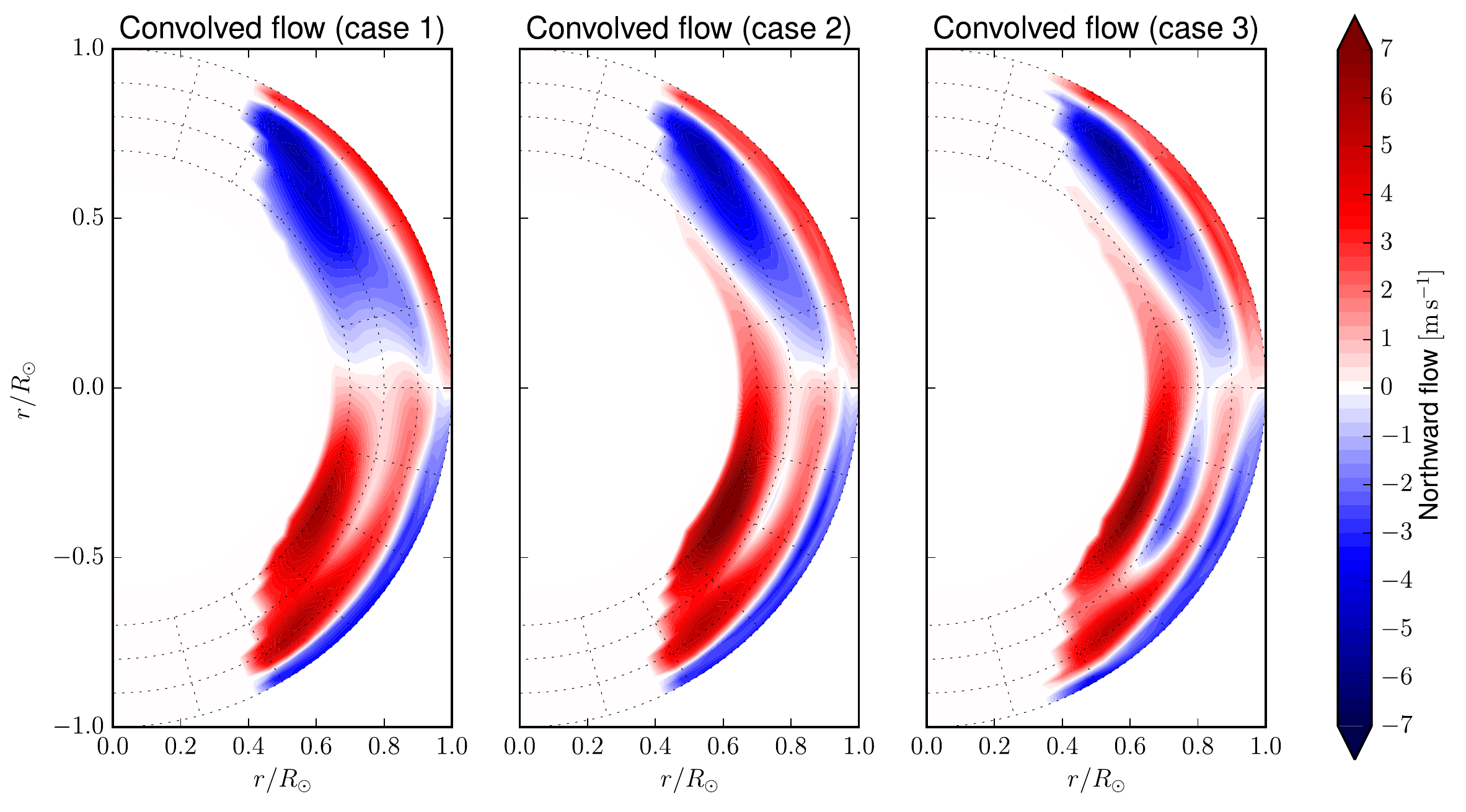}%

\end{center}
\caption{Convolution of the inverted flow profiles from Figure~\ref{fig_comp_flows} with the averaging kernels $\cK_\theta^\theta$. The displayed convolved flows show how the inversion result would look without noise and if the background flow were identical to the inverted flow profile used.}%
\label{fig_comp_vconv}%
\end{figure*}

The possible existence of a multi-cell meridional flow in the southern hemisphere is further underpinned by analyzing the convolution of the inverted flows $v_\theta^{\rm{inv}}$ from Figure~\ref{fig_comp_flows} with the horizontal averaging kernels, that is, Equation~\eqref{eqconvflow} with $m=\theta$. The resulting convolved flow $v_\theta^{\rm{conv}(\theta)}$ is displayed in Figure~\ref{fig_comp_vconv}. It gives an impression as to how the inverted flow would look like if there were no noise and the background flow would be equal to our inversion result $v_\theta^{\rm{inv}}$. As can be seen in Figure~\ref{fig_comp_vconv}, the resulting flow profile would be washed out spatially given the large widths of the averaging kernels in all three cases. Locations with opposite signs near each other are especially no longer seen as clearly. Turning the argument around, it is possible that the locations with a sign change in the inversion result $v_\theta^{\rm{inv}}$ could be even more pronounced in real solar flow.

In case 3, however, the convolved flow looks qualitatively very similar to the inverted flow profile, just with a somewhat lower amplitude. This may lead us to the conclusion that this flow profile is quite a robust result and that the original flow may be similar, just with a somewhat higher amplitude.

\begin{figure*}%

\includegraphics[width=\textwidth]{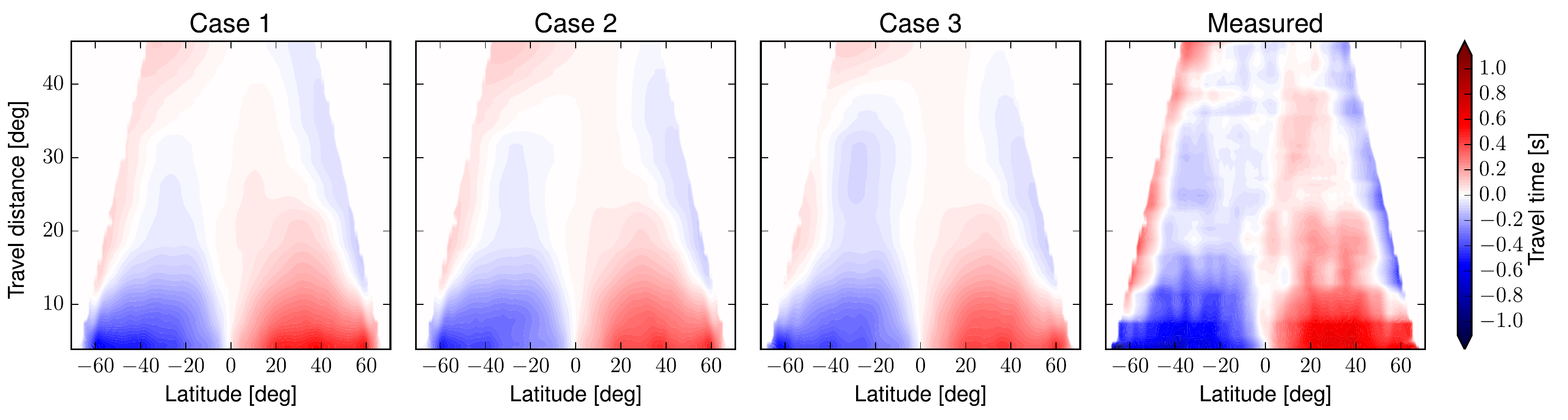}%

\caption{Comparison of forward-modeled travel times obtained from the inversion results from Figure~\ref{fig_comp_flows} to the measured travel times. See also Figure~\ref{figfwdtts1d} for a more detailed comparison of cases 1 and 3 to the measured travel times.}%
\label{fig_comp_tts2d}%
\end{figure*}

\begin{figure*}%
\includegraphics[width=\textwidth]{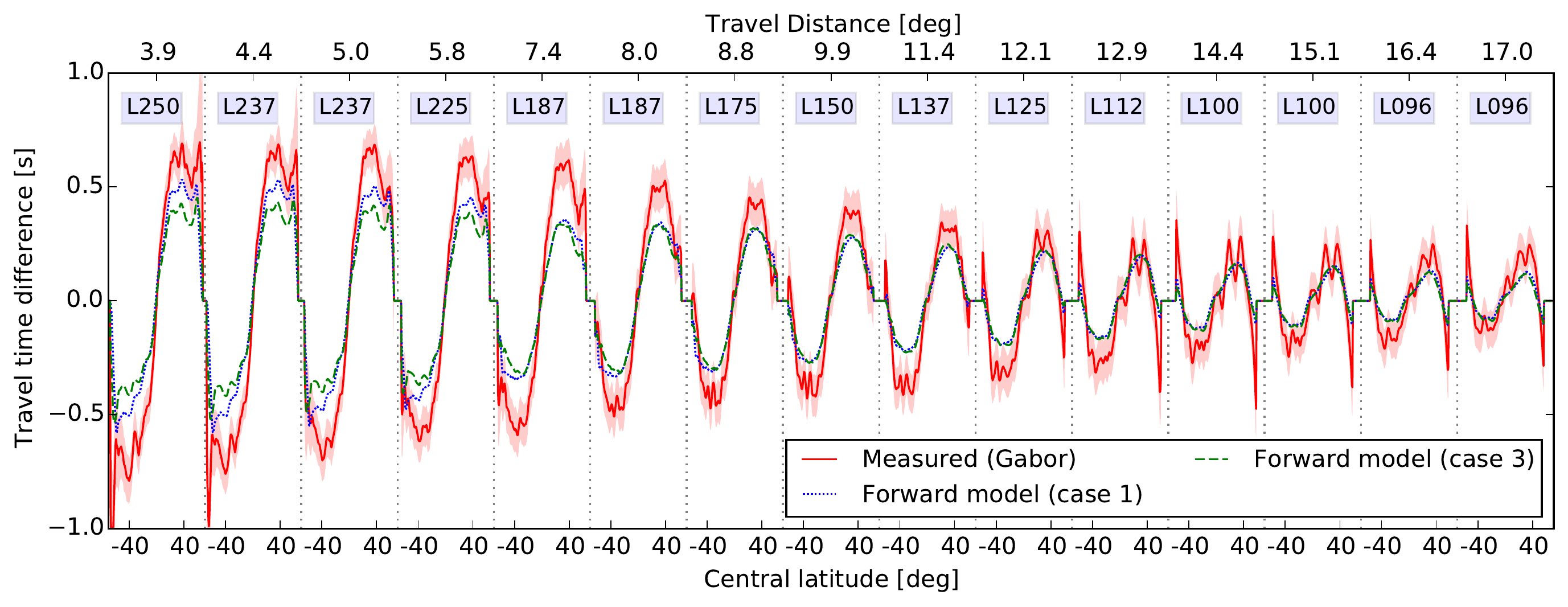}%

\includegraphics[width=\textwidth]{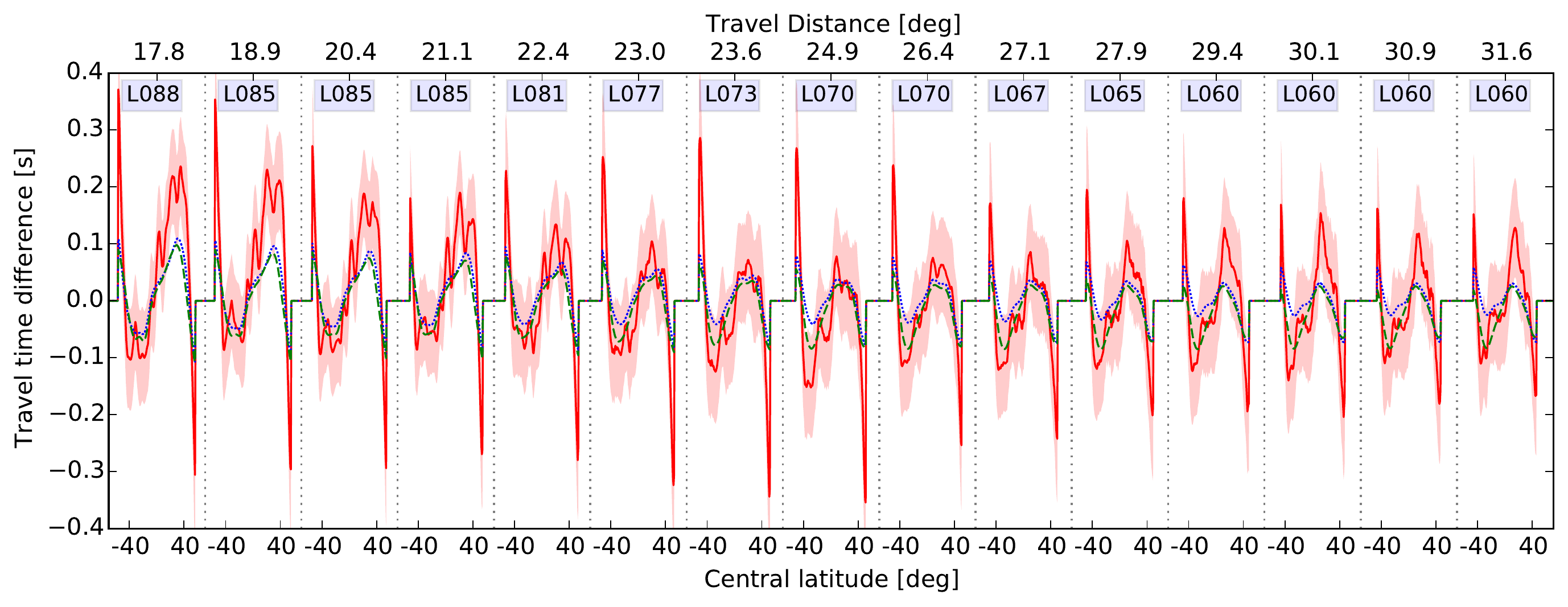}%

\includegraphics[width=\textwidth]{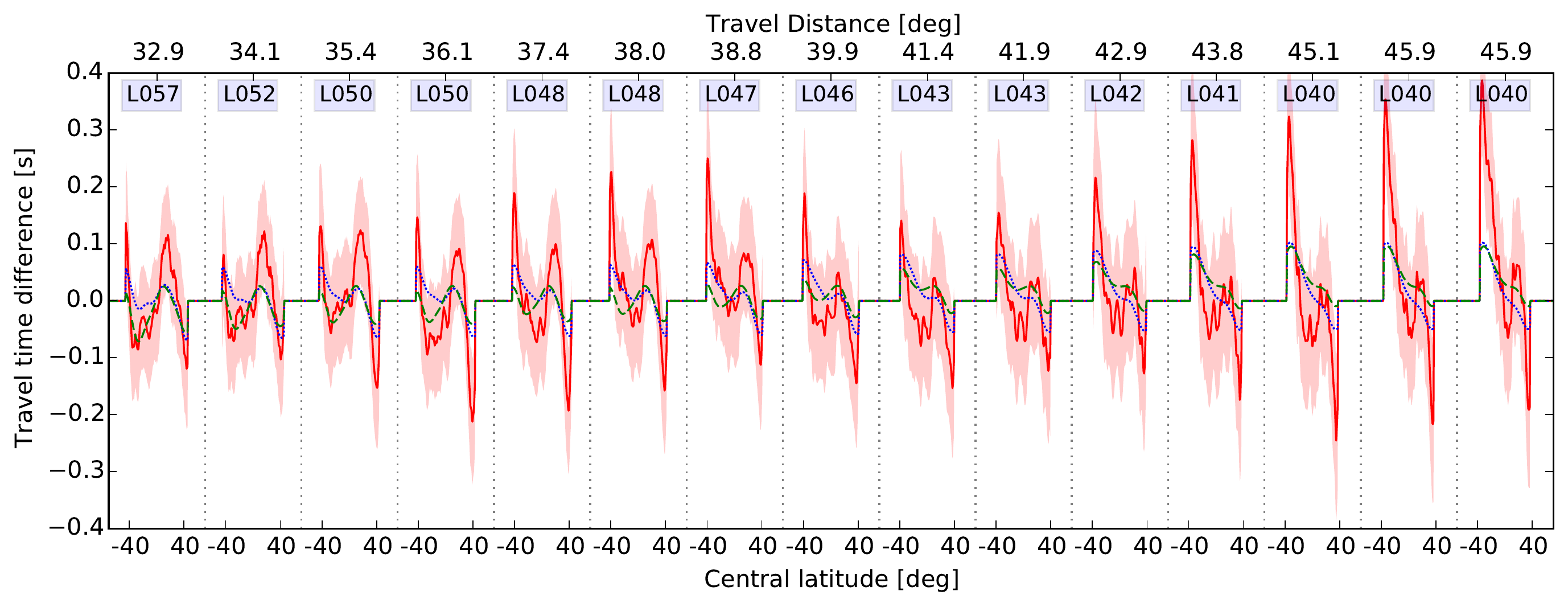}%

\caption{Comparison of measured travel times (solid red) including measurement errors (light red regions) to forward-modeled travel times using the inverted flow of  case 1 (dotted blue) and case 3 (dashed green). All travel times shown in this figure are also shown in Figure~\ref{fig_comp_tts2d}. For all travel distances (upper x-axis), delimited by vertical gray lines, the travel times are plotted as a function of the central latitude of the observation points (bottom x-axis). At each distance, the averaged mean harmonic degree $l$ of the filters applied at each distance is indicated in a blue box.}%
\label{figfwdtts1d}%
\end{figure*}

The match of the inverted flow to the data can be observed in Figures \ref{fig_comp_tts2d} and \ref{figfwdtts1d}, where forward travel times, i.e., the inverted flow convolved with the Born kernels, are compared to the measured travel times. The forward travel times are predominantly consistent with the measured ones within the measurement errors; see Figure~\ref{figfwdtts1d} for cases 1 and 3. We again notice for case 3 that the multi-cell circulation in fact does quite a nice job in reproducing the measured travel times in the southern hemisphere and that a slightly larger magnitude of the flow may explain the measured travel times even better. However, not all features visible in the measured travel times are recovered in the inverted flows, and the errors do not permit us to make a final decision in favor of a single or multiple cell profile of the meridional flow. On the other hand, given the above-mentioned relatively large values for misfit, cross-talk, and the width of the averaging kernels, one could say that the forward-modeled and measured travel times agree remarkably well.

Furthermore, we would like to put forward the somewhat more speculative idea that a multi-cell structure may also do a good job explaining the measured travel times in the northern hemisphere. A hint of an additional cell in the northern hemisphere may be seen in the initial inversion result shown in Figure~\ref{figsolabornall}. Admittedly, such a profile is not seen in the refined inversions of Section~\ref{secfullinv} but we note that it may have been washed out by the large widths of the averaging kernels, especially if the magnitude of the flow is small. 

In addition, we note a few lessons learned in our test inversions, although to some extent of a more incomplete nature. Firstly, it is not clear whether the reversal of signs in the measured travel times at high latitudes -- see travel distances above $9\degr$ and latitudes within about $10\degr$ of the edges in the right panel of Figure~\ref{fig_comp_tts2d} -- is due to a signal from the flow or whether it is caused by a systematic effect. We therefore checked whether excluding this region from the data affects the inversion results for the three cases considered in Section~\ref{secfullinv}. 
The inversion results from the reduced data set are qualitatively very similar to the results shown in Figure~\ref{fig_comp_flows}, especially regarding the general structure and direction of the flow. 
The magnitude of the flow at high latitudes and near the bottom of the convection zone, however, is reduced, which may be seen as physically more realistic regarding mass conservation.

Second, when increasing the threshold for the maximum errors in cases 1 and 2 step by step from $1$ to $2.5\,\mpersec$, the multi-cell structure becomes less pronounced step by step and finally is nearly lost. We obtain a single-cell structure instead. At the same time, the match of the forward-modeled travel times to the measured ones becomes worse. This is not a surprising result, as in this case, the errors reach values similar to the magnitude of the flow in the considered regions. Furthermore, we note that further increasing the threshold for the SVD lets the inversion result further tend toward a smooth single-cell profile, although at some values hints of a second cell may be more pronounced than in case~1.

Finally, we checked that if we do not impose condition~\eqref{eqnormavgKs} with $m=r$ and $k=\theta$, namely, that the integral of the cross-talk kernels be equal to zero, the inversion result does not change much, as the integrals of the cross-talk kernels stay near zero \citep[see also][]{Svanda2011}. Similarly, some tests showed that further decreasing the cross-talk is only possible at the expense of a large increase in the errors and therefore is not an option considered in this work.

\section{Conclusion}
\label{secconclusions}

We have performed SOLA inversions for the deep solar meridional flow using Born kernels. We used GONG data obtained by \cite{Kholikov2014} and inverted by \cite{Jackiewicz2015} using ray kernels. In addition to performing a comparison to the ray kernel inversion, we performed several inversions in order to study systematic effects in the inversion process. 

A first comparison was made using the SOLA inversion technique employed by \cite{Jackiewicz2015}, where the measurement errors were assumed to be uncorrelated (diagonal covariance matrix) and where no regularization for a cross-talk term was done. Our initial inversion result using Born kernels is qualitatively similar to the one obtained using ray kernels by \cite{Jackiewicz2015} while displaying a little more fine structure.

Furthermore, we compared the errors of the inverted flows, which were propagated through the inversion using the diagonal covariance matrix, to the errors computed using the full covariance matrix. Previous results for the deep meridional flow using time-distance helioseismology (\citealp{GilesPhD}, \citealp{Zhao2013,Jackiewicz2015,Rajaguru2015}) were obtained under the assumption of uncorrelated measurements. In our inversion, we find that the errors may be underestimated by a factor of about two to four under such an assumption.

In a subsequent step, we performed refined inversions using the full covariance matrix, thus taking the correlation of the measurements into account, and we introduced a regularization for the cross-talk term as is standard practice in local helioseismology \citep[e.g.,][]{Jackiewicz2008,Svanda2011,DeGrave2014QS}. Three different cases for the choice of inversion parameters were studied. In all of these inversions, we find a shallow return flow at a depth of about $0.9\,R_\sun$, in agreement with \cite{Jackiewicz2015}.

In a first case, we obtained a single-cell meridional flow profile with a few opposite sign ``bumps'', which is again similar to the result obtained by \cite{Jackiewicz2015}, with less latitudinal variation compared to the result obtained with the diagonal covariance matrix. Here, we used a medium threshold for the SVD and about 50\% of the SVs were kept.

When we lower the threshold by a factor of 10 and use about 70\% of the SVs, the refined inversion results exhibit a multi-cell structure in the southern hemisphere and a single-cell meridional circulation structure in the northern hemisphere (cases 2 and 3). Depending on the associated radial flow, the inverted flow has two or three circulation cells stacked radially. At the locations of the additional cells, the inversion result for case 1 already showed a flow structure. This structure was, however, not as pronounced and no conclusion was made on this pattern earlier \citep{Jackiewicz2015}.

In principle, a discussion of the inversion results (see Section~\ref{secdiscussion}) suggests that a multi-cell profile in the southern hemisphere is a suitable candidate as an inversion result, as it is reproduced when convolved with the averaging kernels. All other results are averaged out spatially by this procedure. Furthermore, we speculate that a multi-cell circulation structure may also be present in the northern hemisphere. Indications for this possibility can be seen in the initial result, although it is not present in the refined inversions. However, we point out that both the single-cell and the multi-cell profiles obtained in the refined inversions agree similarly well with the measured travel times within the measurement errors.

Furthermore, the cross-talk of the radial flow into the inversion result for the horizontal flow component was estimated to be below 1.4 $\mpersec$. Therefore, it is only of minor importance, although the cross-talk averaging kernels have relatively large values in all inversions performed in this work.

Finally, a flow structure with multiple cells in latitude as obtained by \cite{Schad2013} using global helioseismology is not seen in the refined inversion results of our study. However, some latitudinal variation can be observed in the travel times, in the initial inversion result, and in the original inversion result from \cite{Jackiewicz2015} using ray kernels. Such a signal may have been averaged out by the relatively wide averaging kernels that have an FWHM of up to $20\degr$. We therefore cannot rule out the possibility of latitudinally stacked flow cells at this point.

Further insights in the inversion problem may be reached in future studies, e.g., by comparing to RLS/LSQR inversion techniques \citep{Zhao2013,Rajaguru2015,Fournier2016}, or to the Pinsker method newly introduced to helioseismology \citep{Fournier2016}. As pointed out by \cite{Fournier2016}, inversions for both radial and horizontal flows are most promising when a mass-conservation constraint is applied; see \cite{Rajaguru2015} for such an inversion. An inversion for radial flows without using mass conservation is not very promising; see \cite{Fournier2016}.

In addition, in order to reach an agreement on the nature of the deep meridional flow, understanding systematic effects such as the center-to-limb effect \citep[e.g.,][]{Zhao2013} and the effect of surface magnetic fields on the measurements \citep[e.g.,][]{Liang2015SurfaceMagnOnMeridFlow} may play a key role in the future.

\acknowledgments

The research leading to these results has received funding from the European Research Council under the European Union's Seventh Framework Programme (FP/2007-2013)/ERC Grant Agreement n. 307117. This work was supported by the SOLARNET project (www.solarnet-east.eu), funded by the European Commission's FP7 Capacities Programme under the Grant Agreement 312495. J.J. acknowledges support from the National Science Foundation under Grant Number 1351311. S.K. was supported by NASA's Heliophysics Grand Challenges Research grant 13-GCR1-2-0036. The authors acknowledge fruitful discussions during the international team meeting on ``Studies of the Deep Solar Meridional Flow'' at ISSI (International Space Science Institute), Bern. This work utilizes data obtained by the Global Oscillation Network Group (GONG) Program, managed by the National Solar Observatory, which is operated by AURA, Inc., under a cooperative agreement with the National Science Foundation. The data were acquired by instruments operated by the Big Bear Solar Observatory, High Altitude Observatory, Learmonth Solar Observatory, Udaipur Solar Observatory, Instituto de Astrofísica de Canarias, and Cerro Tololo Interamerican Observatory. The authors thank Sylvain Korzennik for providing GONG mode frequencies and damping rates. V.B. thanks Damien Fournier for helpful discussions and Juan Manuel Borrero for reading an earlier version of the manuscript. The authors thank the referee for constructive comments that improved the manuscript.




\begin{thebibliography}{}

\bibitem[\protect\astroncite{{Baldner} and {Schou}}{2012}]{Baldner2012}
{Baldner}, C.~S. and {Schou}, J.: 2012,
\newblock {\em \apjl} {\bf 760}, L1,
  {\small[\href{http://adsabs.harvard.edu/abs/2012ApJ...760L...1B}{ADS}]},
\newblock {\small[\href{http://dx.doi.org/10.1088/2041-8205/760/1/L1}{DOI}]}

\bibitem[\protect\astroncite{{Birch} and {Felder}}{2004}]{Birch2004}
{Birch}, A.~C. and {Felder}, G.: 2004,
\newblock {\em \apj} {\bf 616}, 1261,
  {\small[\href{http://adsabs.harvard.edu/abs/2004ApJ...616.1261B}{ADS}]},
\newblock {\small[\href{http://dx.doi.org/10.1086/424961}{DOI}]}

\bibitem[\protect\astroncite{{Birch} and {Gizon}}{2007}]{BG2007}
{Birch}, A.~C. and {Gizon}, L.: 2007,
\newblock {\em AN} {\bf 328}, 228,
  {\small[\href{http://adsabs.harvard.edu/abs/2007AN....328..228B}{ADS}]},
\newblock {\small[\href{http://dx.doi.org/10.1002/asna.200610724}{DOI}]}

\bibitem[\protect\astroncite{{Birch} et~al.}{2004}]{BirchKo2004}
{Birch}, A.~C., {Kosovichev}, A.~G., and {Duvall}, Jr., T.~L.: 2004,
\newblock {\em \apj} {\bf 608}, 580,
  {\small[\href{http://adsabs.harvard.edu/abs/2004ApJ...608..580B}{ADS}]},
\newblock {\small[\href{http://dx.doi.org/10.1086/386361}{DOI}]}

\bibitem[\protect\astroncite{{Birch} et~al.}{2001}]{Birch2001}
{Birch}, A.~C., {Kosovichev}, A.~G., {Price}, G.~H., and {Schlottmann}, R.~B.:
  2001,
\newblock {\em \apjl} {\bf 561}, L229,
  {\small[\href{http://adsabs.harvard.edu/abs/2001ApJ...561L.229B}{ADS}]},
\newblock {\small[\href{http://dx.doi.org/10.1086/324766}{DOI}]}

\bibitem[\protect\astroncite{{B{\"o}ning} et~al.}{2017}]{Boening2017}
{B{\"o}ning}, V.~G.~A., {Roth}, M., {Jackiewicz}, J., and {Kholikov}, S.: 2017,
\newblock {\em \apj} {\bf 838}, 53,
  {\small[\href{http://adsabs.harvard.edu/abs/2017ApJ...838...53B}{ADS}]},
\newblock {\small[\href{http://dx.doi.org/10.3847/1538-4357/aa6333}{DOI}]}

\bibitem[\protect\astroncite{{B{\"o}ning} et~al.}{2016}]{Boening2016}
{B{\"o}ning}, V.~G.~A., {Roth}, M., {Zima}, W., {Birch}, A.~C., and {Gizon},
  L.: 2016,
\newblock {\em \apj} {\bf 824}, 49,
  {\small[\href{http://adsabs.harvard.edu/abs/2016ApJ...824...49B}{ADS}]},
\newblock {\small[\href{http://dx.doi.org/10.3847/0004-637X/824/1/49}{DOI}]}

\bibitem[\protect\astroncite{{Couvidat} et~al.}{2006}]{Couvidat2006}
{Couvidat}, S., {Birch}, A.~C., and {Kosovichev}, A.~G.: 2006,
\newblock {\em \apj} {\bf 640}, 516,
  {\small[\href{http://adsabs.harvard.edu/abs/2006ApJ...640..516C}{ADS}]},
\newblock {\small[\href{http://dx.doi.org/10.1086/500103}{DOI}]}

\bibitem[\protect\astroncite{{DeGrave} and {Jackiewicz}}{2015}]{DeGrave2015}
{DeGrave}, K. and {Jackiewicz}, J.: 2015,
\newblock {\em \solphys} {\bf 290}, 1547,
  {\small[\href{http://adsabs.harvard.edu/abs/2015SoPh..290.1547D}{ADS}]},
\newblock {\small[\href{http://dx.doi.org/10.1007/s11207-015-0693-0}{DOI}]}

\bibitem[\protect\astroncite{{DeGrave} et~al.}{2014a}]{DeGrave2014Sunspot}
{DeGrave}, K., {Jackiewicz}, J., and {Rempel}, M.: 2014a,
\newblock {\em \apj} {\bf 794}, 18,
  {\small[\href{http://adsabs.harvard.edu/abs/2014ApJ...794...18D}{ADS}]},
\newblock {\small[\href{http://dx.doi.org/10.1088/0004-637X/794/1/18}{DOI}]}

\bibitem[\protect\astroncite{{DeGrave} et~al.}{2014b}]{DeGrave2014QS}
{DeGrave}, K., {Jackiewicz}, J., and {Rempel}, M.: 2014b,
\newblock {\em \apj} {\bf 788}, 127,
  {\small[\href{http://adsabs.harvard.edu/abs/2014ApJ...788..127D}{ADS}]},
\newblock {\small[\href{http://dx.doi.org/10.1088/0004-637X/788/2/127}{DOI}]}

\bibitem[\protect\astroncite{{Fournier} et~al.}{2014}]{Fournier2014}
{Fournier}, D., {Gizon}, L., {Hohage}, T., and {Birch}, A.~C.: 2014,
\newblock {\em \aap} {\bf 567}, A137,
  {\small[\href{http://adsabs.harvard.edu/abs/2014A\%26A...567A.137F}{ADS}]},
\newblock {\small[\href{http://dx.doi.org/10.1051/0004-6361/201423580}{DOI}]}

\bibitem[\protect\astroncite{Fournier et~al.}{2016}]{Fournier2016}
Fournier, D., Gizon, L., Holzke, M., and Hohage, T.: 2016,
\newblock {\em Inverse Problems} {\bf 32}, 105002,
  {\small[\href{http://adsabs.harvard.edu/abs/2016InvPr..32j5002F}{ADS}]},
\newblock
  {\small[\href{http://dx.doi.org/10.1088/0266-5611/32/10/105002}{DOI}]}

\bibitem[\protect\astroncite{{Giles}}{2000}]{GilesPhD}
{Giles}, P.~M.: 2000,
\newblock {\em Ph.D. thesis}, Stanford University,
\newblock
  {\small[\href{http://adsabs.harvard.edu/abs/2000PhDT.........9G}{ADS}]}

\bibitem[\protect\astroncite{Gizon et~al.}{2017}]{Gizon2017}
Gizon, L., Barucq, H., Durufl\'{e}, M., Hanson, C.~S., Legu\`{e}be, M., Birch,
  A.~C., Chabassier, J., Fournier, D., Hohage, T., and Papini, E.: 2017,
\newblock {\em Astronomy and Astrophysics} {\bf 600}, A35,
  {\small[\href{http://adsabs.harvard.edu/abs/2017A\%26A...600A..35G}{ADS}]},
\newblock {\small[\href{http://dx.doi.org/10.1051/0004-6361/201629470}{DOI}]}

\bibitem[\protect\astroncite{{Gizon} and {Birch}}{2002}]{GB2002}
{Gizon}, L. and {Birch}, A.~C.: 2002,
\newblock {\em \apj} {\bf 571}, 966,
  {\small[\href{http://adsabs.harvard.edu/abs/2002ApJ...571..966G}{ADS}]},
\newblock {\small[\href{http://dx.doi.org/10.1086/340015}{DOI}]}

\bibitem[\protect\astroncite{{Gizon} and {Birch}}{2004}]{GB2004}
{Gizon}, L. and {Birch}, A.~C.: 2004,
\newblock {\em \apj} {\bf 614}, 472,
  {\small[\href{http://adsabs.harvard.edu/abs/2004ApJ...614..472G}{ADS}]},
\newblock {\small[\href{http://dx.doi.org/10.1086/423367}{DOI}]}

\bibitem[\protect\astroncite{{Gizon} and {Birch}}{2005}]{GB2005}
{Gizon}, L. and {Birch}, A.~C.: 2005,
\newblock {\em Living Rev. Sol. Phys.} {\bf 2}, 6,
  {\small[\href{http://adsabs.harvard.edu/abs/2005LRSP....2....6G}{ADS}]},
\newblock {\small[\href{http://dx.doi.org/10.12942/lrsp-2005-6}{DOI}]}

\bibitem[\protect\astroncite{{Hartlep} et~al.}{2013}]{Hartlep2013}
{Hartlep}, T., {Zhao}, J., {Kosovichev}, A.~G., and {Mansour}, N.~N.: 2013,
\newblock {\em \apj} {\bf 762}, 132,
  {\small[\href{http://adsabs.harvard.edu/abs/2013ApJ...762..132H}{ADS}]},
\newblock {\small[\href{http://dx.doi.org/10.1088/0004-637X/762/2/132}{DOI}]}

\bibitem[\protect\astroncite{Hathaway}{2012}]{Hathaway2012}
Hathaway, D.~H.: 2012,
\newblock {\em ApJ} {\bf 760}, 84,
  {\small[\href{http://adsabs.harvard.edu/abs/2012ApJ...760...84H}{ADS}]},
\newblock {\small[\href{http://dx.doi.org/10.1088/0004-637X/760/1/84}{DOI}]}

\bibitem[\protect\astroncite{{Hazra} et~al.}{2014}]{Hazra2014}
{Hazra}, G., {Karak}, B.~B., and {Choudhuri}, A.~R.: 2014,
\newblock {\em \apj} {\bf 782}, 93,
  {\small[\href{http://adsabs.harvard.edu/abs/2014ApJ...782...93H}{ADS}]},
\newblock {\small[\href{http://dx.doi.org/10.1088/0004-637X/782/2/93}{DOI}]}

\bibitem[\protect\astroncite{{Jackiewicz} et~al.}{2012}]{Jackiewicz2012}
{Jackiewicz}, J., {Birch}, A.~C., {Gizon}, L., {Hanasoge}, S.~M., {Hohage}, T.,
  {Ruffio}, J.-B., and {{\v S}vanda}, M.: 2012,
\newblock {\em \solphys} {\bf 276}, 19,
  {\small[\href{http://adsabs.harvard.edu/abs/2012SoPh..276...19J}{ADS}]},
\newblock {\small[\href{http://dx.doi.org/10.1007/s11207-011-9873-8}{DOI}]}

\bibitem[\protect\astroncite{Jackiewicz et~al.}{2008}]{Jackiewicz2008}
Jackiewicz, J., Gizon, L., and Birch, A.~C.: 2008,
\newblock {\em Solar Physics} {\bf 251}, 381,
  {\small[\href{http://adsabs.harvard.edu/abs/2008SoPh..251..381J}{ADS}]},
\newblock {\small[\href{http://dx.doi.org/10.1007/s11207-008-9158-z}{DOI}]}

\bibitem[\protect\astroncite{{Jackiewicz} et~al.}{2015}]{Jackiewicz2015}
{Jackiewicz}, J., {Serebryanskiy}, A., and {Kholikov}, S.: 2015,
\newblock {\em \apj} {\bf 805}, 133,
  {\small[\href{http://adsabs.harvard.edu/abs/2015ApJ...805..133J}{ADS}]},
\newblock {\small[\href{http://dx.doi.org/10.1088/0004-637X/805/2/133}{DOI}]}

\bibitem[\protect\astroncite{{Kholikov} et~al.}{2014}]{Kholikov2014}
{Kholikov}, S., {Serebryanskiy}, A., and {Jackiewicz}, J.: 2014,
\newblock {\em \apj} {\bf 784}, 145,
  {\small[\href{http://adsabs.harvard.edu/abs/2014ApJ...784..145K}{ADS}]},
\newblock {\small[\href{http://dx.doi.org/10.1088/0004-637X/784/2/145}{DOI}]}

\bibitem[\protect\astroncite{{Korzennik} et~al.}{2013}]{Korzennik2013ApJ}
{Korzennik}, S.~G., {Rabello-Soares}, M.~C., {Schou}, J., and {Larson}, T.~P.:
  2013,
\newblock {\em \apj} {\bf 772}, 87,
  {\small[\href{http://adsabs.harvard.edu/abs/2013ApJ...772...87K}{ADS}]},
\newblock {\small[\href{http://dx.doi.org/10.1088/0004-637X/772/2/87}{DOI}]}

\bibitem[\protect\astroncite{{Kosovichev} and
  {Duvall}}{1997}]{KosovichevDuvall1997}
{Kosovichev}, A.~G. and {Duvall}, Jr., T.~L.: 1997,
\newblock in F.~P. {Pijpers}, J. {Christensen-Dalsgaard}, and C.~S. {Rosenthal}
  (eds.), {\em SCORe'96 : Solar Convection and Oscillations and their
  Relationship}, Vol. 225 of {\em Astrophysics and Space Science Library}, pp
  241--260,
\newblock
  {\small[\href{http://adsabs.harvard.edu/abs/1997ASSL..225..241K}{ADS}]},
\newblock {\small[\href{http://dx.doi.org/10.1007/978-94-011-5167-2_26}{DOI}]}

\bibitem[\protect\astroncite{Liang et~al.}{2017}]{Liang2017}
Liang, Z.-C., Birch, A.~C., Duvall, T.~L., Gizon, L., and Schou, J.: 2017,
\newblock {\em Astronomy \& Astrophysics} {\bf 601}, A46,
  {\small[\href{http://adsabs.harvard.edu/abs/2017arXiv170400475L}{ADS}]},
\newblock {\small[\href{http://dx.doi.org/10.1051/0004-6361/201730416}{DOI}]}

\bibitem[\protect\astroncite{{Liang} and
  {Chou}}{2015a}]{Liang2015SurfaceMagnOnMeridFlow}
{Liang}, Z.-C. and {Chou}, D.-Y.: 2015a,
\newblock {\em \apj} {\bf 805}, 165,
  {\small[\href{http://adsabs.harvard.edu/abs/2015ApJ...805..165L}{ADS}]},
\newblock {\small[\href{http://dx.doi.org/10.1088/0004-637X/805/2/165}{DOI}]}

\bibitem[\protect\astroncite{{Liang} and
  {Chou}}{2015b}]{Liang2015ProbingMagneticFields}
{Liang}, Z.-C. and {Chou}, D.-Y.: 2015b,
\newblock {\em \apj} {\bf 809}, 150,
  {\small[\href{http://adsabs.harvard.edu/abs/2015ApJ...809..150L}{ADS}]},
\newblock {\small[\href{http://dx.doi.org/10.1088/0004-637X/809/2/150}{DOI}]}

\bibitem[\protect\astroncite{{Miesch}}{2005}]{Miesch2005LRSP}
{Miesch}, M.~S.: 2005,
\newblock {\em Living Rev. Sol. Phys.} {\bf 2}, 1,
  {\small[\href{http://adsabs.harvard.edu/abs/2005LRSP....2....1M}{ADS}]},
\newblock {\small[\href{http://dx.doi.org/10.12942/lrsp-2005-1}{DOI}]}

\bibitem[\protect\astroncite{Pijpers and Thompson}{1994}]{Pijpers1994}
Pijpers, F.~P. and Thompson, M.~J.: 1994,
\newblock {\em Astronomy and Astrophysics} {\bf 281}, 231,
  {\small[\href{http://adsabs.harvard.edu/abs/1994A\%26A...281..231P}{ADS}]}

\bibitem[\protect\astroncite{{Rajaguru} and {Antia}}{2015}]{Rajaguru2015}
{Rajaguru}, S.~P. and {Antia}, H.~M.: 2015,
\newblock {\em \apj} {\bf 813}, 114,
  {\small[\href{http://adsabs.harvard.edu/abs/2015ApJ...813..114R}{ADS}]},
\newblock {\small[\href{http://dx.doi.org/10.1088/0004-637X/813/2/114}{DOI}]}

\bibitem[\protect\astroncite{{Roth} et~al.}{2016}]{Roth2016}
{Roth}, M., {Doerr}, H.-P., and {Hartlep}, T.: 2016,
\newblock {\em \aap} {\bf 592}, A106,
  {\small[\href{http://adsabs.harvard.edu/abs/2016A\%26A...592A.106R}{ADS}]},
\newblock {\small[\href{http://dx.doi.org/10.1051/0004-6361/201526971}{DOI}]}

\bibitem[\protect\astroncite{{Roth} et~al.}{2007}]{Roth2007}
{Roth}, M., {Gizon}, L., and {Beck}, J.~G.: 2007,
\newblock {\em AN} {\bf 328}, 215,
  {\small[\href{http://adsabs.harvard.edu/abs/2007AN....328..215R}{ADS}]},
\newblock {\small[\href{http://dx.doi.org/10.1002/asna.200610722}{DOI}]}

\bibitem[\protect\astroncite{{Schad} et~al.}{2013}]{Schad2013}
{Schad}, A., {Timmer}, J., and {Roth}, M.: 2013,
\newblock {\em \apjl} {\bf 778}, L38,
  {\small[\href{http://adsabs.harvard.edu/abs/2013ApJ...778L..38S}{ADS}]},
\newblock {\small[\href{http://dx.doi.org/10.1088/2041-8205/778/2/L38}{DOI}]}

\bibitem[\protect\astroncite{{SILSO World Data
  Center}}{2012}]{SSN_SILSO_2004_2012}
{SILSO World Data Center}: 2004-2012,
\newblock {\em International Sunspot Number Monthly Bulletin and online
  catalogue},
\newblock {\small[\href{http://www.sidc.be/silso/}{URL}]}

\bibitem[\protect\astroncite{{{\v S}vanda}}{2013}]{Svanda2013a}
{{\v S}vanda}, M.: 2013,
\newblock {\em \apj} {\bf 775}, 7,
  {\small[\href{http://adsabs.harvard.edu/abs/2013ApJ...775....7S}{ADS}]},
\newblock {\small[\href{http://dx.doi.org/10.1088/0004-637X/775/1/7}{DOI}]}

\bibitem[\protect\astroncite{{{\v S}vanda} et~al.}{2011}]{Svanda2011}
{{\v S}vanda}, M., {Gizon}, L., {Hanasoge}, S.~M., and {Ustyugov}, S.~D.: 2011,
\newblock {\em \aap} {\bf 530}, A148,
  {\small[\href{http://adsabs.harvard.edu/abs/2011A\%26A...530A.148S}{ADS}]},
\newblock {\small[\href{http://dx.doi.org/10.1051/0004-6361/201016426}{DOI}]}

\bibitem[\protect\astroncite{{{\v S}vanda}
  et~al.}{2013a}]{Svanda2013Validation}
{{\v S}vanda}, M., {Roudier}, T., {Rieutord}, M., {Burston}, R., and {Gizon},
  L.: 2013a,
\newblock {\em \apj} {\bf 771}, 32,
  {\small[\href{http://adsabs.harvard.edu/abs/2013ApJ...771...32S}{ADS}]},
\newblock {\small[\href{http://dx.doi.org/10.1088/0004-637X/771/1/32}{DOI}]}

\bibitem[\protect\astroncite{{{\v S}vanda} et~al.}{2013b}]{Svanda2013c}
{{\v S}vanda}, M., {Schunker}, H., and {Burston}, R.: 2013b,
\newblock {\em Journal of Physics Conference Series} {\bf 440(1)}, 012024,
  {\small[\href{http://adsabs.harvard.edu/abs/2013JPhCS.440a2024S}{ADS}]},
\newblock
  {\small[\href{http://dx.doi.org/10.1088/1742-6596/440/1/012024}{DOI}]}

\bibitem[\protect\astroncite{{Zhao} et~al.}{2013}]{Zhao2013}
{Zhao}, J., {Bogart}, R.~S., {Kosovichev}, A.~G., {Duvall}, Jr., T.~L., and
  {Hartlep}, T.: 2013,
\newblock {\em \apjl} {\bf 774}, L29,
  {\small[\href{http://adsabs.harvard.edu/abs/2013ApJ...774L..29Z}{ADS}]},
\newblock {\small[\href{http://dx.doi.org/10.1088/2041-8205/774/2/L29}{DOI}]}

\bibitem[\protect\astroncite{{Zhao} et~al.}{2012a}]{Zhao2012}
{Zhao}, J., {Couvidat}, S., {Bogart}, R.~S., {Parchevsky}, K.~V., {Birch},
  A.~C., {Duvall}, T.~L., {Beck}, J.~G., {Kosovichev}, A.~G., and {Scherrer},
  P.~H.: 2012a,
\newblock {\em \solphys} {\bf 275}, 375,
  {\small[\href{http://adsabs.harvard.edu/abs/2012SoPh..275..375Z}{ADS}]},
\newblock {\small[\href{http://dx.doi.org/10.1007/s11207-011-9757-y}{DOI}]}

\bibitem[\protect\astroncite{{Zhao} et~al.}{2012b}]{Zhao2012CtoL}
{Zhao}, J., {Nagashima}, K., {Bogart}, R.~S., {Kosovichev}, A.~G., and
  {Duvall}, Jr., T.~L.: 2012b,
\newblock {\em \apjl} {\bf 749}, L5,
  {\small[\href{http://adsabs.harvard.edu/abs/2012ApJ...749L...5Z}{ADS}]},
\newblock {\small[\href{http://dx.doi.org/10.1088/2041-8205/749/1/L5}{DOI}]}

\bibitem[\protect\astroncite{Zhao et~al.}{2016}]{Zhao2016}
Zhao, J., Stejko, A., and Chen, R.: 2016,
\newblock {\em Solar Physics} {\bf 291}, 731,
  {\small[\href{http://adsabs.harvard.edu/abs/2016SoPh..291..731Z}{ADS}]},
\newblock {\small[\href{http://dx.doi.org/10.1007/s11207-016-0864-7}{DOI}]}

\end{thebibliography}

%
%

\end{document}